\documentclass{aa}

\usepackage{graphicx}
\usepackage{txfonts}
\usepackage{upgreek}    
\begin{document}

   \title{KIC 8553788: A pulsating Algol with an extreme mass ratio}

   \author{A. Liakos
           }
   \institute{Institute for Astronomy, Astrophysics, Space Applications and Remote Sensing, National Observatory of Athens,\\
            Metaxa \& Vas. Pavlou St., GR-15236, Penteli, Athens, Greece\\\\
              \email{alliakos@noa.gr}
             }

   \date{Received September XX, 2017; accepted March XX, 2017}


\abstract
   {The present research paper focuses on the eclipsing binary KIC~8553788 which belongs to two different types of binary systems regarding its physical properties. In particular, it is one of the 71 oscillating stars of Algol type that have been discovered so far and one of the 6 that have been published based on high cadence photometric data of the $Kepler$ mission. In addition, it is one of the four semi-detached binaries of the group of R~CMa type systems, while its pulsating component has the fourth fastest frequency among the $\delta$~Scuti stars-members of semi-detached binaries. Detailed light curves, spectroscopic and pulsation analyses are presented, while possible explanation scenarios for the evolution of the system involving past mass transfer, mass loss and/or angular momentum loss due to the presence of a tertiary component are discussed.}
   {The goal of the study is to extract the pulsational characteristics of the oscillating star of the system, to estimate the absolute parameters of its components and to provide possible explanation for its extreme evolutionary status.}
   {Ground-based spectroscopic observations using the 2.3~m ``Aristarchos'' telescope were obtained and used for the estimation of the spectral type of the primary component and to model the light curves of the system with higher certainty. The short cadence photometric data provided by the $Kepler$ mission were analysed using standard eclipsing binary modelling techniques, while Fourier analysis was applied on their residuals aiming to reveal the properties of the intrinsic oscillations. The resulted photometric model was combined with published radial velocity curve to obtain accurate absolute parameters for the components of the system.}
   {The results show that the primary component of the system is of A8 spectral type, has a mass of 1.6~$M_{\sun}$, and a radius of 2~$R_{\sun}$. It is a relatively fast pulsator of $\delta$~Scuti type that oscillates in 89 frequency modes with the dominant one to be 58.26~cycles~day$^{-1}$. On the other hand, the secondary component has a mass of only 0.07~$M_{\sun}$, a radius of 1~$R_{\sun}$, and a temperature of 4400~K. In addition, it was found to be magnetically active with migrating cool spots on its surface.}
   {KIC~8553788 according to its geometrical configuration and its pulsational properties belongs to the group of the oscillating stars of Algol type, while according to its very low mass ratio and its relatively short orbital period belongs also to the group of R~CMa stars. If confirmed by radial velocity data of the secondary component, the system would have the lowest mass ratio that has ever been found in semi-detached systems and it can be considered as one of the most extreme cases.}


   \maketitle
%

\section{Introduction}
\label{sec:intro}

Generally, eclipsing binary systems (hereafter EBs) can be considered as the ultimate tools for the calculation of stellar absolute parameters that are needed to check the current stellar evolutionary models. Photometric and spectroscopic observations of EBs can be used to derive directly stellar masses, radii, luminosities, etc. In particular, the light curves (hereafter LCs) analysis provides the means to derive the geometrical configuration (e.g. detached, semi-detached), the inclination, the orbital period etc. of the system. Moreover, short to med-time scale (i.e. from days to a few years) photometric observations offer the opportunity to check for the existence of physical mechanisms that occur in the components (e.g. magnetic activity, intrinsic variability). Another valuable photometric application is the eclipse timing variations (ETV) method that provide us with information regarding long-time scale (i.e. many years) physical mechanisms (e.g. mass transfer, additional components) that also contribute to the evolution of the system. On the other hand, spectroscopy is another key-method to obtain accurate results, especially when measuring the radial velocities (hereafter RVs) of both components. Although it is very often the spectra of systems with large temperature difference between their components to be dominated by the light of the brighter star, it is feasible to proceed in the absolute parameters determination using fair assumptions based on the stellar evolution theory.

The oscillating stars of $\delta$~Scuti type are multiperiodic variables pulsating in radial and non-radial modes driven mostly by the $\kappa$-mechanism \citep{BAL15} in the frequency range 4-80~cycle~d$^{-1}$ \citep{BRE00}. They are located in the classical instability strip, range between A-F spectral types and III-V luminosity classes, and have masses between 1.4-3~$M_{\sun}$.

Asteroseismic studies of pulsating stars-members of EBs are quite important mainly because using the binarity as a tool is feasible to calculate the absolute parameters of the pulsator, which is essential for asteroseismic evolutionary models. In particular, $\delta$~Scuti stars in binaries have become targets of high scientific interest during the last two decades, especially after the availability of high quality data from space missions. Although some cases were known from the 70's, the interest for this kind of systems was not that great, probably because of the low quality data that led to ambiguous results. However, after 2000 there has been an increased growth of interest in these systems, that led \citet{MKR02} to propose them as a separate group of EBs. They defined them as the classical Algol-type systems with mass accreting pulsators of (B)A-F spectral type, commonly known as oEA~stars (oscillating eclipsing binaries of Algol type). Systematic observations aiming to discover such systems increased their number and, later, led many researchers to focus on their properties \citep[e.g. Orbital-pulsational periods, $\log g$ - pulsational period correlations,][]{SOY06a, LIA12, ZHA13, LIAN17, KAH17}. The most complete catalogue (204 $\delta$~Scuti stars in 199 binaries) for these systems to date, which also includes updated correlations between their fundamental parameters, was published by \citet{LIAN17} and is also available in an online form\footnote{\url{http://alexiosliakos.weebly.com/catalogue.html}}. It should to be noted that 9 more systems of this type have been discovered after that publication with the total number of systems increased to 208.

\begin{figure}
\centering
\includegraphics[width=\columnwidth]{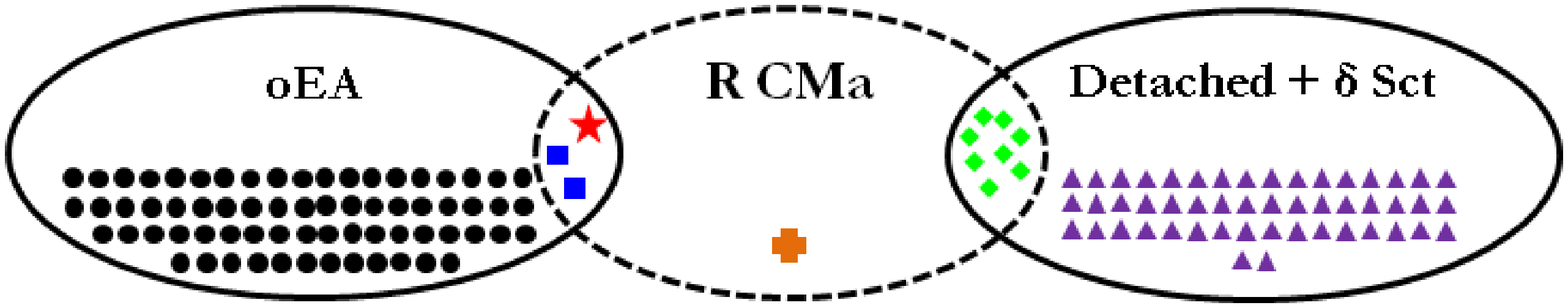}
\caption{Overlap between the groups of oEA stars (71 members, denoted by dots), detached binaries with a $\delta$~Scuti component (58 members, denoted by triangles), and R~CMa systems. Squares and diamonds represent the systems that belong to two different groups and the cross symbol the system OGLEGC~228, which is the only R~CMa system that does not exhibit pulsations. The red star denotes KIC~8553788.}
\label{fig:groups}
\includegraphics[width=\columnwidth]{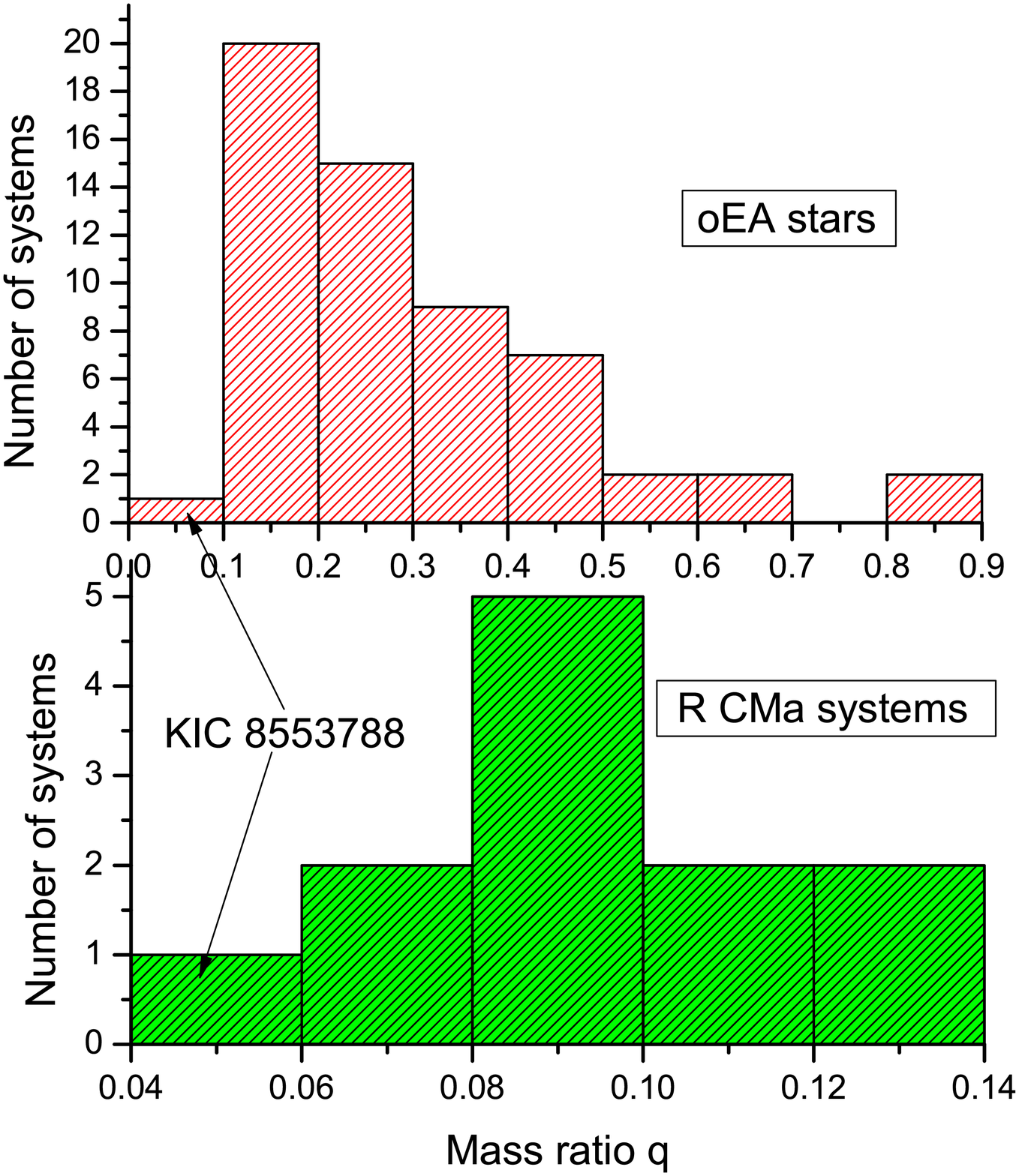}
\caption{The mass ratio distribution for oEA stars (upper section) and R~CMa systems (lower section). The position of KIC~8553788 is also indicated.}
\label{fig:qdist}
\end{figure}

The R~CMa systems were introduced by \citet{BUD11} as the classical Algol EBs with unusual combination of very low mass ratio and relatively short orbital period. The members of this group are only 11 so far, namely R~CMa \citep{BUD11}, OO~Dra \citep{LEE18}, AS~Eri \citep{MKR04, IBA06}, KIC~8087799 \citep{ZHA17}, KIC~8262223 \citep{GUO17}, KIC~9285587 \citep{FAI15}, KIC~10661783 \citep{LEH13}, KIC~10989032 \citep{ZHA17}, KIC~11401845 \citep{LEE17}, OGLEGC~228 \citep{KAL07}, and WASP~1628+10 \citep{MAX14}. R~CMa and AS~Eri are also oEA stars, while KIC~8087799, KIC~8262223, KIC~9285587, and WASP~1628+10 are detached systems with a $\delta$~Scuti and an He white dwarf (WD) components. Therefore, it should to be noted that all R~CMa systems, except for OGLEGC~228, include a pulsating component of $\delta$~Scuti type. Fig.~\ref{fig:groups} shows the overlap between the groups of binaries hosting a $\delta$~Scuti component (oEA and detached systems) and the R~CMa systems.

The $Kepler$ space mission can be plausibly considered as the most valuable data treasure for asteroseismic studies. In particular, the high accuracy of the measurements ($\sim10^{-4}$~mag), the continuous monitoring of  targets for several days, and the time resolution ($\sim1$~min) of the short cadence (SC) data provide the means for very detailed pulsation modelling. In addition, the monitoring of a target during the various quarters of the mission offers the opportunity to study its behavior on med-time scale range (i.e. few years). Especially for the EBs, it should be emphasized the excellent online database, namely $Kepler$ Eclipsing Binary Catalog\footnote{\url{http://keplerebs.villanova.edu/}} \citep[KEBC,][]{PRS11}, that includes all the preliminary information and the data for several thousands binaries.

KIC~8553788 (=2MASS~J19174291+4438290, $\alpha_{2000}=19^{\rm h}~17^{\rm m}~42.914^{\rm s},~\delta_{2000}=+44\degr~38'~29.08''$) was discovered as an EB with an orbital period of $\sim$1.606~days by the ASAS project \citep[ASAS J191743+4438.5,][]{PIG09}. It was included in the long cadence data of all quarters (i.e. Q0-Q17) of the $Kepler$ mission between 2009-2013 and during the quarters Q5 and Q14 was observed also in SC mode. \citet{GIE12} mentioned for the first time the possibility of pulsations and starspots existence in the system and their affect on the minima timings. Additionally, they published the ETV analysis of the system proposing the Light-travel effect due to a tertiary component as the most realistic solution for the cyclic changes of its orbital period. Later on, \citet{GIE15}, \citet{ZAS15}, and \citet{BOR16} confirmed that result and determined the period of the third body as 8.6, 9.1, and 12.5~yrs, respectively. \citet{FRA16}, based on two spectra obtained by the LAMOST survey, classified the system between A5IV-A8III spectral types, while \citet{MAT17} included it in their spectroscopic survey on $Kepler$ EBs and published the RV of the primary component.

The motivation for this work is based on the rarity of the nature of KIC~8553788. On one hand, only another five oEA systems observed by the $Kepler$ mission have been announced so far \citep{LIAN17, LEE16, LIA17} (the total known are 70) and only three semi-detached systems of R~CMa type are known to date (see Fig.~\ref{fig:groups}). On the other hand, KIC~8553788 has potentially the smallest mass ratio that has ever been found (see Fig.~\ref{fig:qdist}, for details see Section~\ref{sec:LCmdl}) in semi-detached systems of these kinds, whose mass ratio is known. Its primary component exhibits pulsations, while its secondary has the smallest mass value ever found among such systems and, in general, in semi-detached systems. Undoubtedly, this EB can be considered as one of the most extreme cases of binaries hosting a pulsating component.

\begin{figure*}[h!]
\centering
\includegraphics[width=18cm]{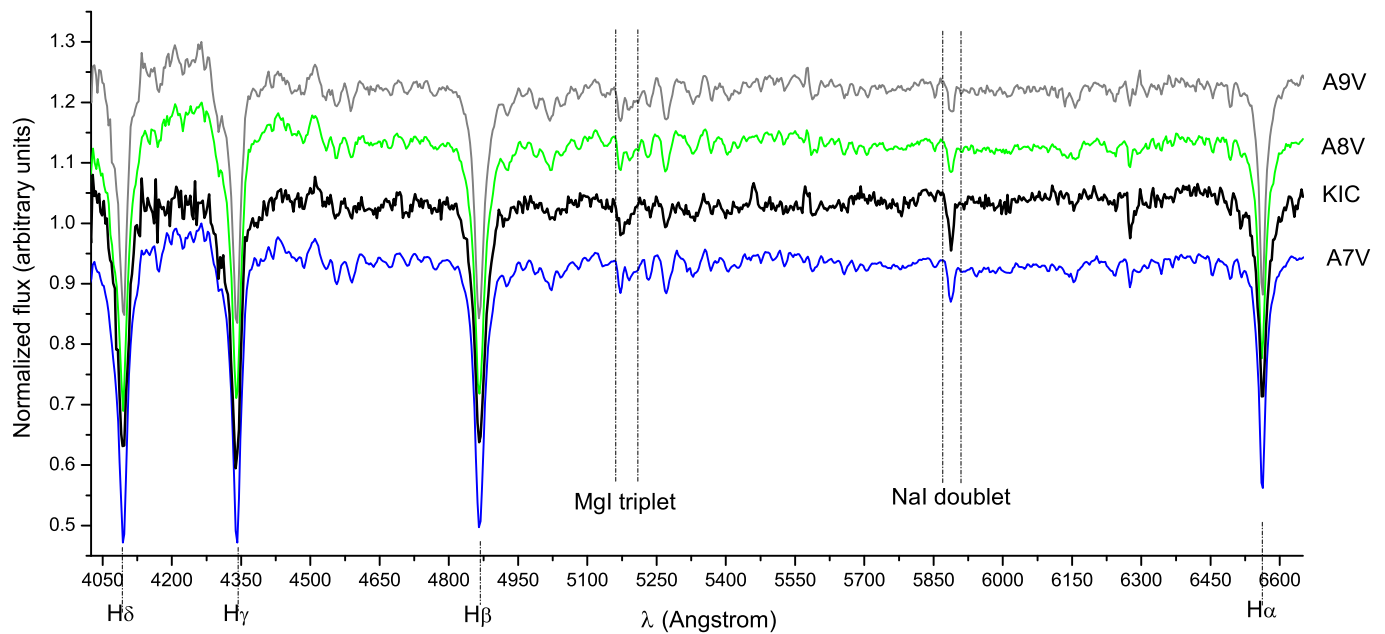}
\caption{Comparison spectra of KIC~8553788 and standard stars with the closest spectral types. The Balmer and some strong metallic lines are also indicated. The A7V spectrum belongs to HIP~13904, the A8V to HIP~90345, and the A9V to HIP~10830.}
\label{fig:spectra}
\end{figure*}

Here, a detailed analysis of the SC $Kepler$ LCs of KIC~8553788 will be presented. Based on the published RV \citep{MAT17} and ground-based spectroscopic observations, it is feasible for the first time to obtain an accurate photometric model of the system and calculate the absolute parameters of its components with a relatively high precision. Then, a detailed asteroseismic analysis on the light curve residuals follows the LC solution, while, additionally, the variation of magnetic spots is also presented. Finally, the physical properties of the system are compared with those of other systems of both oEA and R~CMa type and its extraordinary evolutionary status evolving mass transfer, mass loss and angular momentum redistribution scenarios is discussed.

\section{Ground-based Spectroscopy}
\label{sec:sp}

Spectroscopic observations were made in 2016 with the 2.3~m Ritchey-Cretien ``Aristarchos'' telescope at Helmos Observatory in Greece using the Aristarchos Transient Spectrometer\footnote{\url{http://helmos.astro.noa.gr/ats.html}} \citep[ATS,][]{BOU04}. The ATS is a low-to-medium dispersion fibre spectrometer that consists of 50 fibres (50~$\mu$m diameter each) providing a field-of-view of $\sim10$~arcsec on the sky and is equipped with the U47-MB Apogee (Back illuminated, 1024$\times$1024~pixels, 13~$\mu$m$^{2}$~pixel size) CCD camera. The low resolution grating (600~lines~mm$^{-1}$) was used for the observations, which provided a spectral coverage between approximately 4000~{\AA}-7260~{\AA} and a resolution of $\sim3.2$~{\AA}~pixel$^{-1}$.

The observations were aimed towards the spectral classification of the primary component of KIC~8553788 in order to use it later for the photometric modelling (see section~\ref{sec:LCmdl}). For this, approximately 45 spectroscopic standard stars, suggested by GEMINI Observatory\footnote{\url{http://www.gemini.edu/}}, ranging from B0V to K8V spectral types were observed between July-October with the same instrumental set-up as that of the target star. The system was observed on 6 Oct 2016 during the orbital phase $\sim0.31$. Two spectra with an integration time of 15~min were obtained and stacked in order to achieve a better signal-to-noise ratio (S/N). The pre-reduction of the images containing science and calibration spectra, which includes bias, dark, and flat-field corrections, was made using the \textsc{MaxIm DL} software, while the data reduction (wavelength calibration,
cosmic rays removal, spectra normalization, sky background removal) was performed using the \textsc{RaVeRe} v.2.2c software \citep{NEL09}.

All spectra were normalized in order to be directly comparable and then shifted, using the Balmer lines as reference, to compensate for the relative Doppler shifts of each standard and the system under study. The Balmer (i.e. H$_{\alpha}$-H$_{\delta}$) and many strong metallic lines (e.g. Mg$_{\rm I}$-triplet, Na$_{\rm I}$-doublet), that are sensitive to the temperature, were used for the spectral classification. In particular, the depths of the aforementioned lines were calculated in each spectrum of the standard stars and compared with those of KIC~8553788 deriving sums of squared residuals ($\sum \rm res^{2}$) in each case. The least squares sum guided us to the closest match between the spectra of target and standards.

\begin{figure}
\centering
\includegraphics[width=\columnwidth]{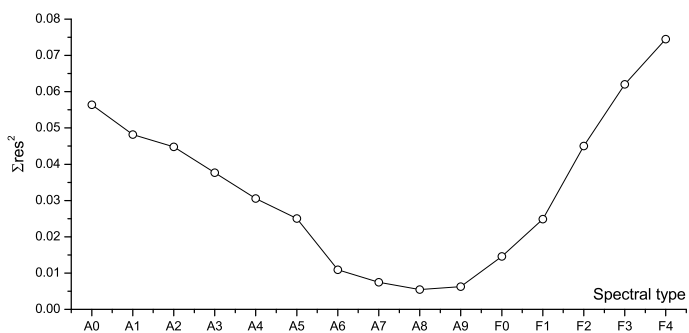}
\caption{Spectral type-search plot for KIC~8553788. Points denote the differences between the spectrum of the system and those of the standard stars of various spectral types. The comparison is shown only between A-F spectral types due to scaling reasons.}
\label{fig:STs}
\end{figure}

The best match of the spectrum of the system was found with that of an A8V standard star. Comparison spectra of KIC~8553788 and standard stars with the closest spectral types are illustrated in Fig.~\ref{fig:spectra}. The respective spectral type-search plot that shows the differences in terms of $\sum \rm res^{2}$ between the spectra of the system under study and the standard stars is given in Fig.~\ref{fig:STs}. Although the spectra of KIC~8553788 were obtained during an out of eclipse phase, it can be plausibly assumed that they reflect the spectral type of its primary component (i.e. more luminous), since, as it will be shown in the next section, the secondary contributes less than 4$\%$ to the total luminosity.

Concluding, according to the above method and the effective temperature-spectral type relation of \citet{COX00} the primary component of KIC~8553788 is an A8V star with a formal error of one subclass and is assigned a temperature of 7600$\pm200$~K. This result comes in slight disagreement with the temperature given in the KEBC catalogue \citep[8045~K,][photometry based method]{PRS11}, in very good agreement with the spectral type given in the Tycho catalogue \citep[A7V,][color index based method]{PIC10}, and is quite close to the classification based on the LAMOST data \citep[A5IV-A8III,][spectroscopic method]{FRA16}.

\section{Light curve modelling and absolute parameters calculation}
\label{sec:LCmdl}
KIC~8553788 was included in all long-cadence quarters of the $Kepler$ mission, but only in Q5 and Q14 the observations were made in SC mode. Given that two of the main goals of the present paper are (a) the very accurate LC modelling, that will lead to (b) a high precision asteroseismic modelling, only the SC data \citep[taken from KEBC,][]{PRS11} were used. The SC data offer continuous recording of a target (i.e. no time gaps) with a relatively high time resolution and provide the means to detect both slow and fast pulsation frequencies. In particular, 46154 and 34832 points that correspond to Q5 and Q14 data sets, respectively, were used to obtain the LC models. The observations cover 19 full LCs during Q5 and 14 during Q14, so, approximately 2350 points per LC are available. It should to be noted, that according to the MAST archive the level of light contamination of the system is zero.

The first step for the LC modelling was the creation of one simple mean LC in order to apply the ``$q$-search'' method and estimate roughly the photometric mass ratio of the system, since only one RV curve was available (see Section~\ref{sec:intro}). The main problem was that the LCs are superimposed by the pulsations features, that, in general, affect negatively the modelling, while, additionally, photospheric spots produce extra asymmetries. In order to overcome the aforementioned inhomogeneities, all the LCs were folded into the EB's orbital period and average points (i.e. approximately 1 mean point every 135 original points) were calculated. The final binned LC contained approximately only 600 data points. The ephemeris used for the phase folding is given in the KEBC \citep{PRS11} and is included also in Table~\ref{tab:LCmdlAbs}.

The second step was the rough estimation of the photometric mass ratio of the system. For this, the software \textsc{PHOEBE} v.0.29d \citep{PRS05}, that is based on the 2003 version of the Wilson-Devinney (W-D) code \citep{WIL71, WIL79, WIL90}, was used. In absence of spectroscopic mass ratio, the ``$q$-search'' method \citep[for details see][]{LIAN12} was applied in modes 2 (detached system), 4 (semi-detached system with the primary component filling its Roche lobe) and 5 (conventional semi-detached binary). The step of $q$ change was 0.05-0.1 starting from $q=0.025$. The effective temperature of the primary was given the present spectroscopic value (see Section~\ref{sec:sp}) and the third light option was initially enabled, given that the system may have a tertiary component (see Section~\ref{sec:intro}). The method followed for the LC fitting regarding the fixed (filter, albedos $A$, gravity $g$, and limb darkening $x$ coefficients) and the adjusted parameters (inclination $i$, Roche potentials $\Omega$, relative luminosities, temperature of the secondary component) is described in detail in \citet{LIA17}.

\begin{figure}
\includegraphics[width=\columnwidth]{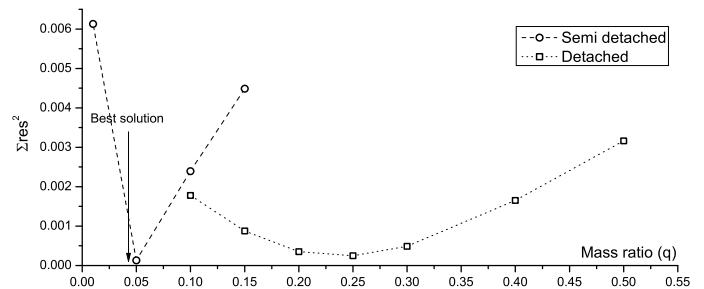}
\caption{$q$-search plots for KIC~8553788 in detached (squares) and semi-detached modes (circles).}
\label{fig:qs}
\end{figure}

The ``$q$-search'' method showed very good results for both modes 2 and 5. The difference in terms of $\sum \rm res^{2}$ is small, as shown in the $q$-search plot in Fig.~\ref{fig:qs}, with the solution of mode 5 to be marginally better. In particular, the minimum $\sum \rm res^{2}$ was found for $q=0.05$ in mode 5 and for $q=0.25$ in mode 2. Initially, it seemed that the mean LC of this system could be described sufficiently by both modes, so, potentially, we would not be able to determine the properties of its components with a unique solution. Fortunately, the answer for this dilemma was given by the RV of the primary component \citep[$K_{1}=8.6\pm0.7$~km~s$^{-1}$,][]{MAT17}. If the model of mode 2 was adopted, then the derived absolute parameters of the components would result in very unrealistic values (i.e. the masses derived are $M_{1}=0.012~M_{\sun}$ and $M_{2}=0.003~M_{\sun}$). Therefore, mode 5 was selected as the most appropriate solution. It should be noted that mode 2 was also tried to fit the LC using very low $q$ values (i.e. 0.02-0.1), following the works of \citet{LEH13} and \citet{LEE17, LEE18}, but after each iteration the secondary component was always overflowing its Roche lobe. The latter was also tested with and without the use of the third light parameter but the results did not show any significant difference.

\begin{figure}
\begin{tabular}{c}
\includegraphics[width=\columnwidth]{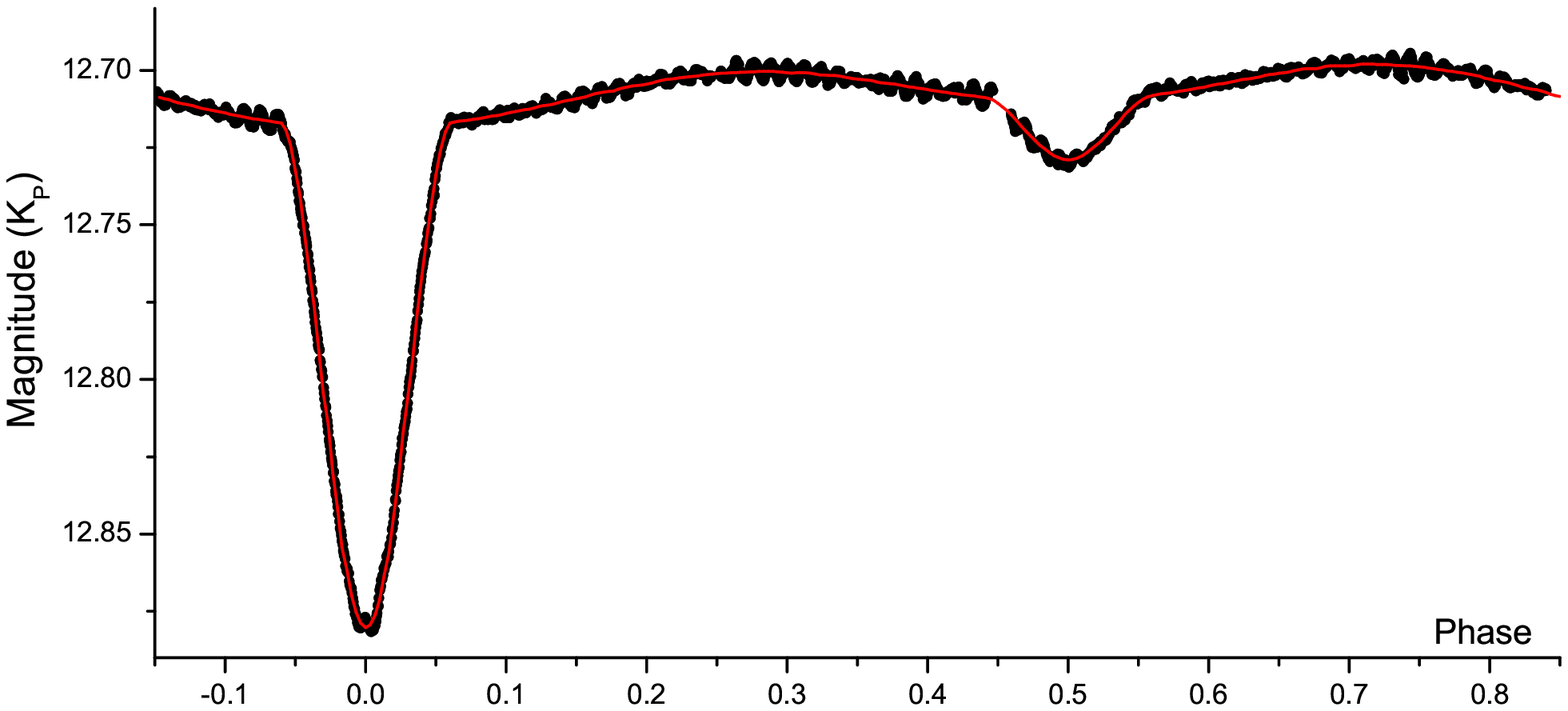}\\
\includegraphics[width=6cm]{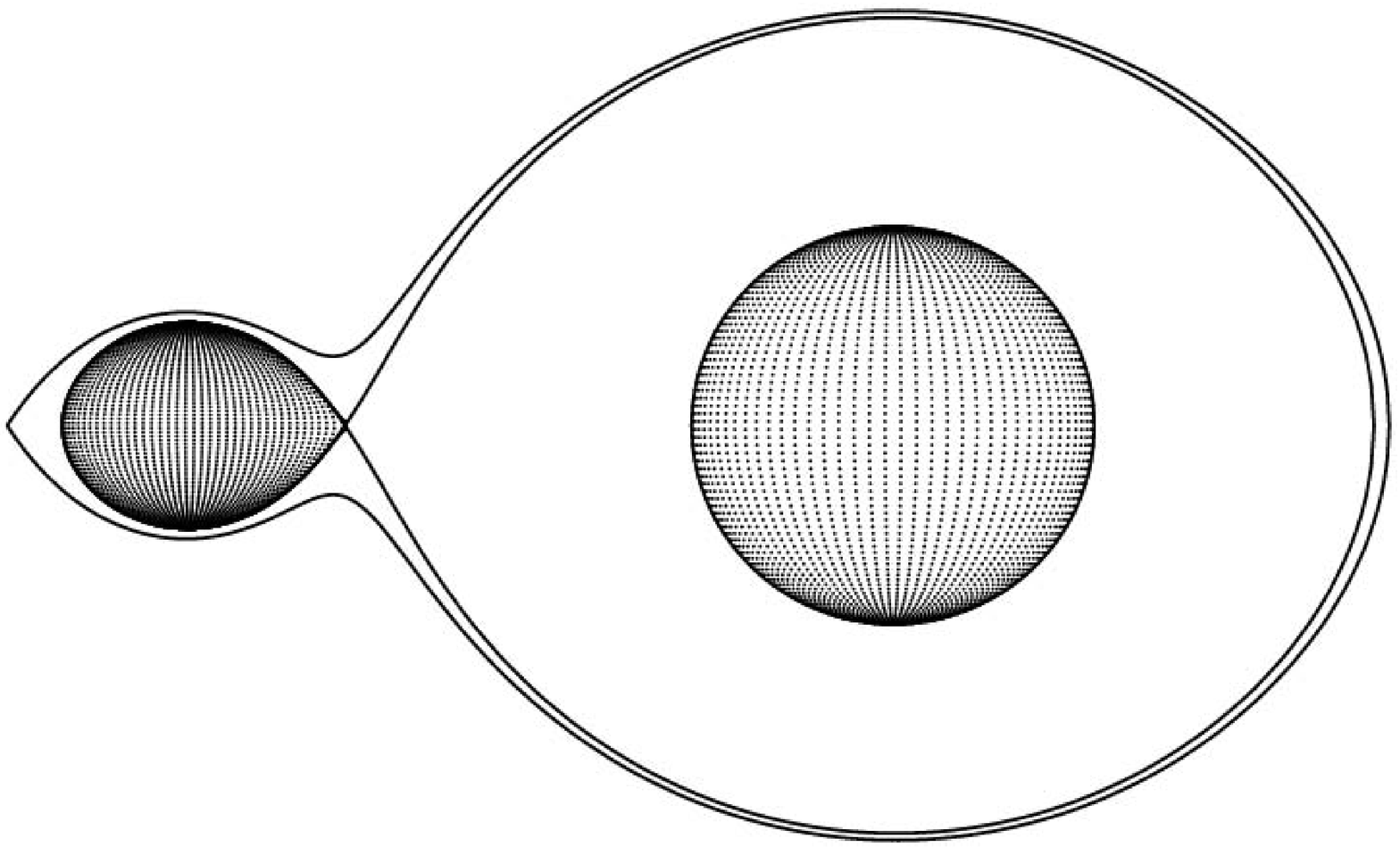}
\end{tabular}
\caption{Upper panel: Theoretical (solid line) over observed (points) LC for KIC~8553788. Lower panel: Three-dimensional representation of the Roche geometry of the system at orbital phase 0.75.}
\label{fig:LCm3D}
\end{figure}

\begin{figure*}
\centering
\begin{tabular}{c}
\includegraphics[width=16cm]{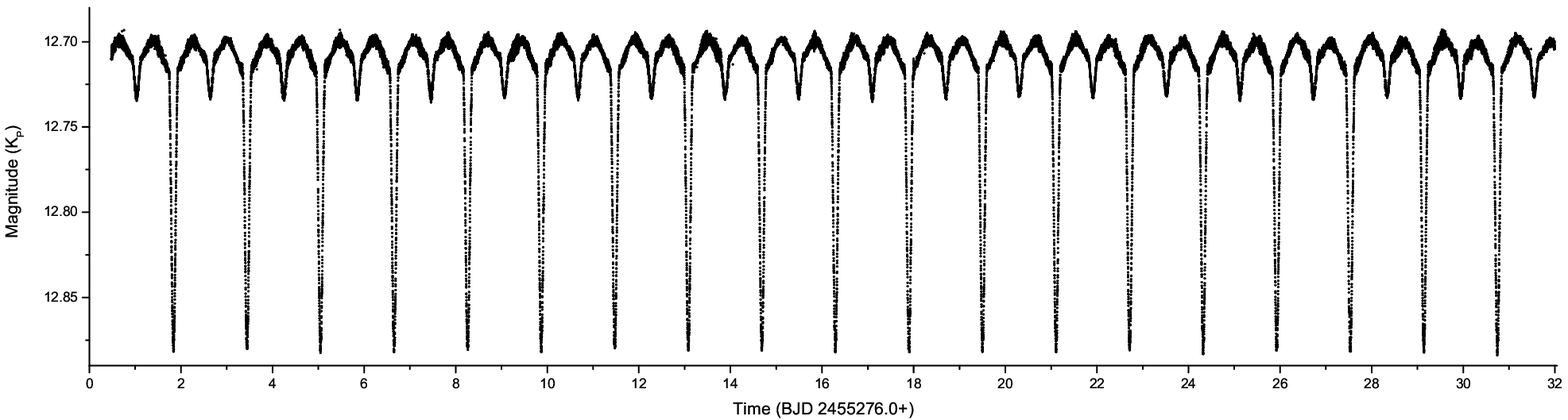}\\
\includegraphics[width=16cm]{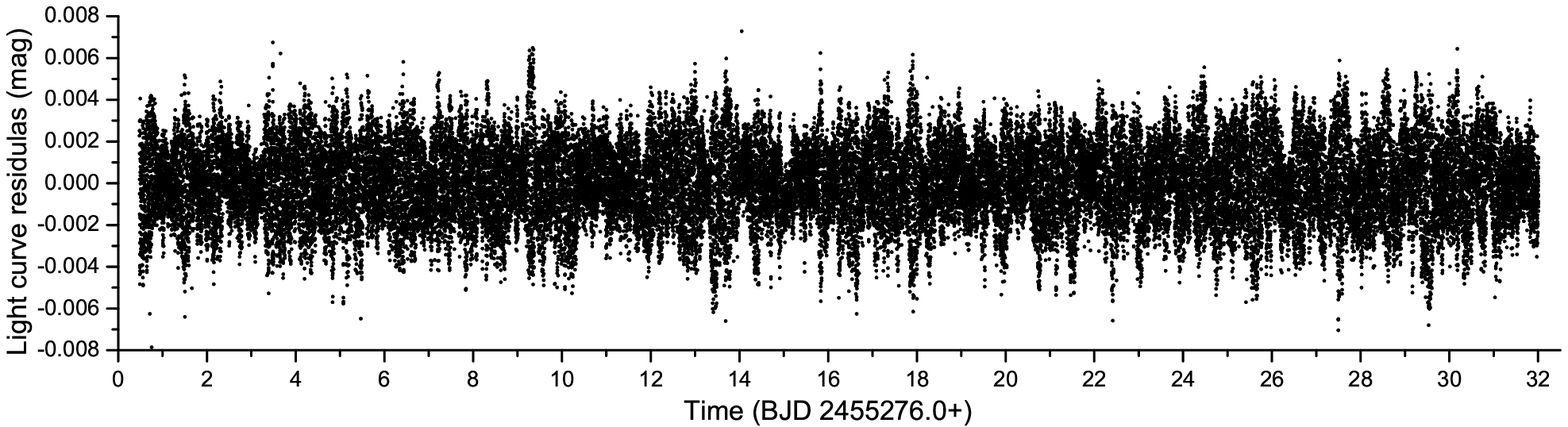}\\
\includegraphics[width=16cm]{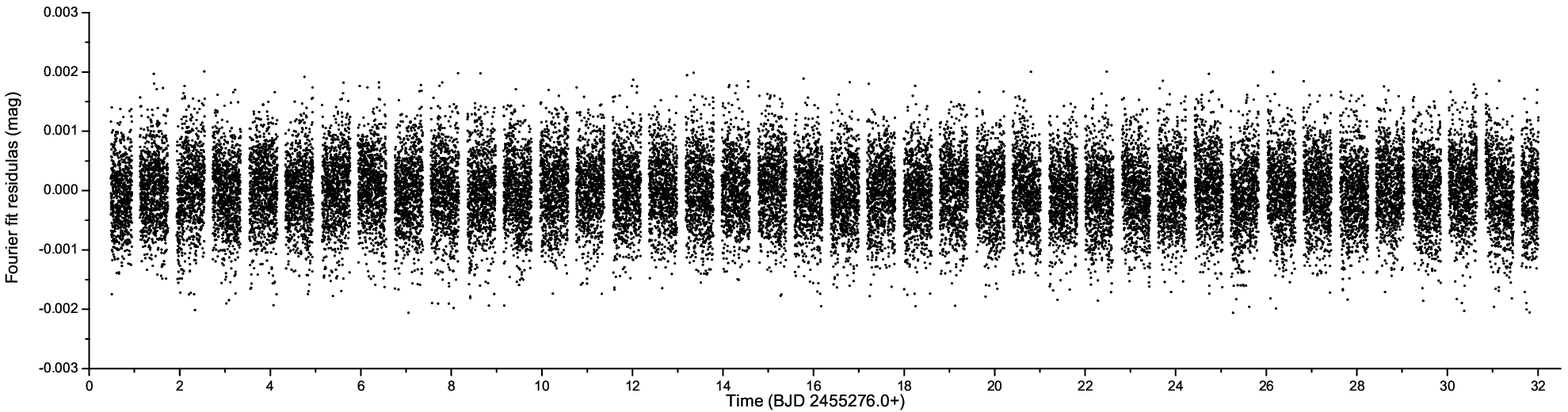}\\
\end{tabular}
\caption{Upper panel: Short cadence LCs of Q5 for KIC~8553788. Middle panel: LCs residuals after the subtraction of the binary models. Bottom panel: LCs residuals after the subtraction of the pulsation model.}
\label{fig:LCs1}
\end{figure*}

\begin{figure*}
\centering
\begin{tabular}{c}
\includegraphics[width=16cm]{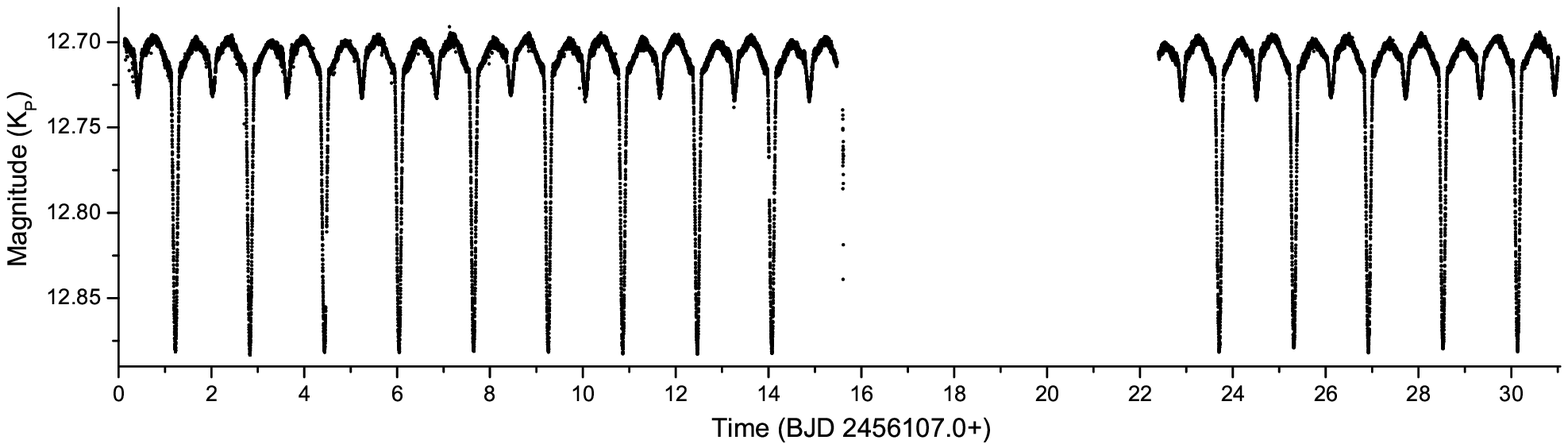}\\
\includegraphics[width=16cm]{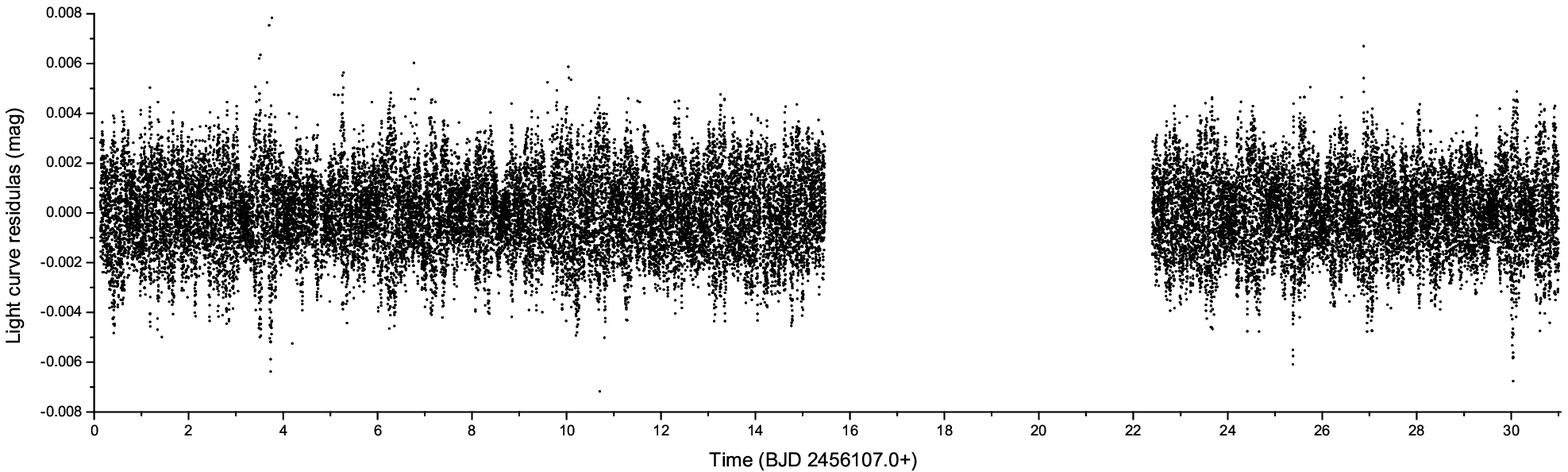}\\
\end{tabular}
\caption{Upper panel: Short cadence LCs of Q14 for KIC~8553788. Bottom panel: LCs residuals after the subtraction of the binary models.}
\label{fig:LCs2}
\end{figure*}

\begin{table}
\centering
\caption{Light curve and absolute parameters for KIC~8553788.}
\label{tab:LCmdlAbs}
\begin{tabular}{l cc }
\hline					
	&  	\multicolumn{2}{c}{System parameters}			\\
\hline					
$K_{\rm p}^{\rm a}$~(mag)	&	\multicolumn{2}{c}{12.691}			\\
$T_{0}^{\rm a}$~(BJD)	&	\multicolumn{2}{c}{2454954.998(30) }			\\
$P_{\rm orb}^{\rm a}$~(days)	&	\multicolumn{2}{c}{1.606163(2) }			\\
$q~(M_{2}/M_{1})$ 	&	\multicolumn{2}{c}{0.043(3) }			\\
$i~(\degr)$	&	\multicolumn{2}{c}{74.84(5) }			\\
\hline					
	&	\multicolumn{2}{c}{Components parameters}			\\
\hline					
        	&	Primary	&	Secondary	\\
\hline					
$T_{\rm eff}$~(K)	&	7600(200)$^{\rm b}$	&	4399(110)	\\
$\Omega$                  	&	3.583(9)	&	1.760(3)	\\
$A^{\rm c}$ 	&	1	&	0.5	\\
$g^{\rm c}$ 	&	1	&	0.32	\\
$x$	&	0.427	&	0.707	\\
$L/(L_{\rm P}+L_{\rm S})$ 	&	0.966(1)	&	0.034(2)	\\
$r_{\rm pole}$	&	0.282(1)	&	0.146(1)	\\
$r_{\rm point}$	&	0.286(1)	&	0.221(1)	\\
$r_{\rm side}$	&	0.286(1)	&	0.152(1)	\\
$r_{\rm back}$	&	0.286(1)	&	0.180(1)	\\
\hline					
	&	\multicolumn{2}{c}{Absolute parameters}			\\
\hline					
$M~$($M_{\sun})$	&	1.6(5)	&	0.07(1)	\\
$R~$($R_{\sun})$	&	2.0(2)	&	1.0(1)	\\
$L~$($L_{\sun})$	&	11(3)	&	0.3(1)	\\
$\log g~$(cm~s$^{-2}$)	&	4.1(2)	&	3.3(1)	\\
$a$~($R_{\sun})$	&	0.28(2)	&	6.6(7)	\\
$M_{\rm bol}$~(mag)       	&	2.1(3)	&	5.9(9)	\\
$K$~(km~s$^{-1}$)	&	8.6(7)$^{\rm d}$	&	200(21)	\\
\hline					
\end{tabular}					
\newline					
\textbf{Notes.} The errors are given in parentheses alongside values and correspond to the last digit(s). $^{\rm (a)}$Taken from the KEBC \citep{PRS11}, $^{\rm (b)}$Result from spectroscopy (Section~\ref{sec:sp}), $^{\rm (c)}$assumed, $^{\rm (d)}$taken from \citet{MAT17}					
\end{table}	
The final model of each individual LC in mode 5 was achieved by setting $q$=0.05 as initial mass ratio value, which was adjusted later during the iterations. Similarly, the rest parameters were initially given values that derived from the best solution of the ``$q$-search'' method and then were left free during the iterations. Due to maxima brightness changes between successive LCs, one photospheric spot on the surface of the secondary component was assumed for the data of Q5, while two spots on the same star were needed for the Q14 data set. The selection of the magnetically active component was based on the temperature values. The primary star is of A8 spectral type (see Section~\ref{sec:sp}), hence, it is very unlikely to host a convective envelope, that is required for photospheric spots, in contrast with the secondary, which is much cooler. Moreover, as it will be shown in the next section, the primary is a $\delta$~Sct pulsator and therefore magnetic activity exhibition is quite unlikely. The third light parameter was also adjusted, but since it resulted in unrealistic values, it was omitted from the final solutions.

In Table~\ref{tab:LCmdlAbs} are presented three sets of parameters: (a) The system parameters, (b) the component parameters, and (c) the absolute parameters of the components. The (a) and (b) sets contain both assumed and derived values for all the parameters used for the LCs models. However, since 19 and 14 models for Q5 and Q14 data sets, respectively, were derived, the values of the adjusted parameters are the averages of the same parameters of the individual LC models, while the errors are the standard deviation of them, except for the $T_2$ and $q$ quantities, which are discussed below. In order to achieve a more realistic error for $T_2$, the error of $T_1$ was taken into account. Tests on individual LCs showed that $T_2$ is very sensitive to $T_1$ changes, and specifically that for 200~K decrease or increase of $T_1$ (i.e. the spectroscopic error of $T_1$), $T_2$ value follows a respective change of approximately 100~K. Therefore, this error was summed in quadrature to the standard deviation of $T_2$ as derived from the LC fittings. Moreover, the error of mass ratio is another matter of discussion. In particular, the value derived as the standard deviation of the respective $q$ values from all the LC models was 0.001, which seems too low when compared with spectroscopic determined mass ratio errors \citep[c.f.][]{MAT17}. This obviously is correlated with the density of the data points available for an RV fit (i.e. few dozens at the best) and the respective data points density of a photometric LC (i.e. order of several thousands). However, in order to obtain a more realistic error value for $q$, the mass function is taken into account using the $P_{\rm orb}$ and $K_1$ values and their errors:
\begin{equation}
f(m)=\frac{M^{3}_{2} \sin^{3} i}{(M_{1}+M_{2})^{2}}=\frac{P_{\rm orb} K^{3}_{1}}{2\pi G},
\end{equation}
where $M_1$, $M_2$, $K_1$, $i$, and $P_{\rm orb}$ are explained in Table~\ref{tab:LCmdlAbs} and $G$ is the gravitational constant. Following the error propagation method and assuming $M_1=1.6~M_{\sun}$, then according to the definition of mass ratio (Table~\ref{tab:LCmdlAbs}), the error of $q$ is derived as 0.0031, which is adopted as the final value. The parameters of the set (c) were calculated using the software \textsc{AbsParEB} \citep{LIA15} in mode 2 and they have the usual meaning. In Fig.~\ref{fig:LCm3D} is shown the photometric model on one LC and the 3D Roche geometry representation of the system, while in Fig.~\ref{fig:LCs1} (upper and middle panels) and Fig.~\ref{fig:LCs2} are illustrated the SC data and their residuals after the subtraction of the theoretical LCs for the two data sets.

During the modelling of the individual LCs, it was noticed that the spots migrate from one day to another. For this, although this is a by product result and a detailed study is not an aim of the present paper, their parameters are listed in Table~\ref{tab:spots} in Appendix~\ref{sec:App2}. Moreover, in the same appendix are given plots showing the location changes of the spots as well as their location on the surface of the secondary component on the first and the last days of each data set (Figs~\ref{fig:spotmigr1}-\ref{fig:spotmigr2b}).

The reliability of the solution, especially for the extremely low value of the mass ratio, may be checked in the future using the RVs of the secondary component. However, it should to be noted that according to \citet[][sec. 4.1.1.1]{KAL99} the photometric mass ratio can be determined with much confidence for lobe-filling or overcontact systems when at least one light and one RV curves exist. Moreover, another useful and recent example regarding the reliability of the ``$q$-search'' method concerns the semidetached systems KIC~10581918 and KIC~10619109. \citet{LIA17} determined their mass ratio values based only on their Kepler LCs and without any use of RVs. These photometric values of $q$ were later come in very close agreement with the spectroscopically determined ones of \citet{MAT17}.

Concluding, KIC~8553788 was found to be in semi-detached state with the less massive and cooler component filling its Roche lobe and has a very low mass ratio value of 0.043. Its primary component is a dwarf star, while its secondary, although much less massive, is more evolved, and, additionally, presents magnetic activity. Further discussion for the evolution of this EB is given in Section~\ref{sec:Evol}.

\section{Pulsations analysis}
\label{sec:Fmdl}
Except for the light variations due to successive eclipses between the components of the system and the magnetic activity of its secondary, more periodic variations of much shorter time scales were found to occur. This is obvious also in the upper section of Fig.~\ref{fig:LCm3D}, especially in the phase parts between the eclipses. The short term variability is clearly attributed to intrinsic pulsations.

In order to identify these oscillations, frequency analysis was made with the software \textsc{PERIOD04} v.1.2 \citep{LEN05}, that is based on classical Fourier analysis, on the LCs residuals. The typical frequencies range for $\delta$~Sct stars is between 4-80~cycle~d$^{-1}$ \citep{BRE00}, thus the analysis should be made for this regime. Moreover, taking into account that in many $\delta$~Sct stars in binary systems $g$-mode pulsations, that are connected to their $P_{\rm orb}$, may occur or even though these stars may be $\gamma$~Dor-$\delta$~Sct hybrids, the search was extended to 0-80~cycle~d$^{-1}$.

Although the SC data of Q14 covers relatively enough observational time ($\sim31$~days), they present a long time gap of approximately 7 days between the 16th and 22nd days of observations. Given the critical role of continuous data points in the frequency search \citep[e.g. false detections of alias frequencies,][]{BRE00}, the data of Q14 were not included in the analysis. On the other hand, the data of Q5 do not suffer from any gaps, they cover a $\sim32$~days time period, therefore, they serve much better the purposes of such an analysis. The LCs residuals derived from the subtraction of photometric model from the observed data points (see Section~\ref{sec:LCmdl}) are considered unaffected by any binary influences (e.g. eclipses, asymmetries due to spots). However, although the secondary component has a very small light contribution to the total light of the EB, it was found wise to exclude the data obtained during the eclipses (i.e. between 0.445-0.555 and 0.940-1.060 phase parts), since the pulsations that occur in these phase parts are affected with a non uniform way, in contrast with those in the rest regions (i.e. constant light from both components).

In order to calculate all the significant pulsation frequencies, the same method for the background noise estimation as stated in \citet{LIA17} was followed. In particular, the background noise was calculated in the range 70-80~cycle~d$^{-1}$, where no obvious frequencies exist, a spacing of 2~cycle~d$^{-1}$ and a box size of 2 were used and it was found to be 9.18~$\mu$mag. Regarding the significance of a detected frequency, the same limit (i.e. 4$\sigma$, S/N$>4$), as that suggested by the software was adopted. Therefore the threshold for a validated frequency was $3.67\times10^{-2}$~mmag. The Nyquist number of the data set was 725.02 and the frequency resolution according to the Rayleigh-Criterion (i.e. 1.5/$T$, where $T$ is the observations time range in days) was 0.0469~cycle~d$^{-1}$. Finally, after the first frequency computation, the residuals were subsequently pre-whitened for the next one until the detected frequency had S/N$\sim$4.

\begin{figure}
\includegraphics[width=8.5cm]{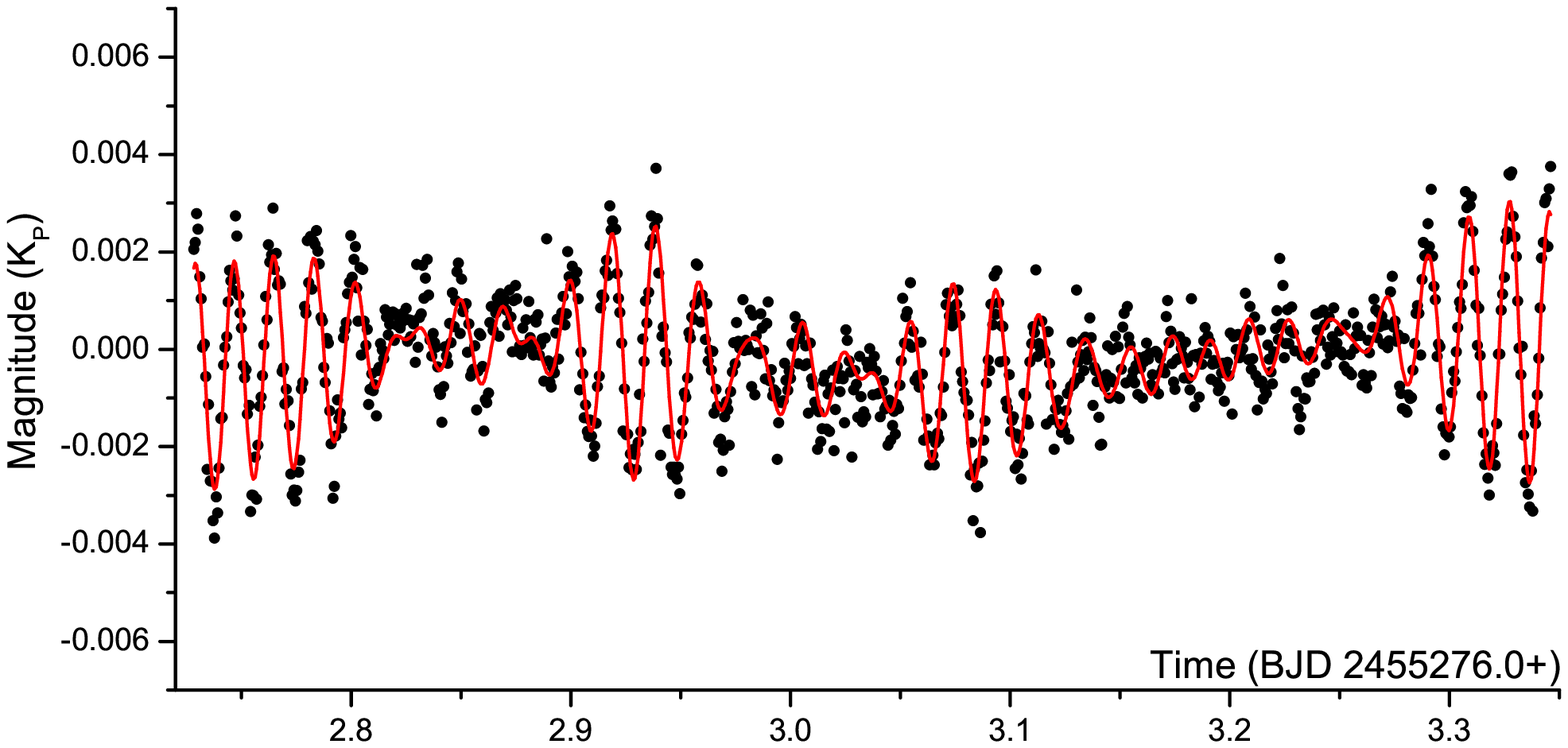}\\
\includegraphics[width=8.5cm]{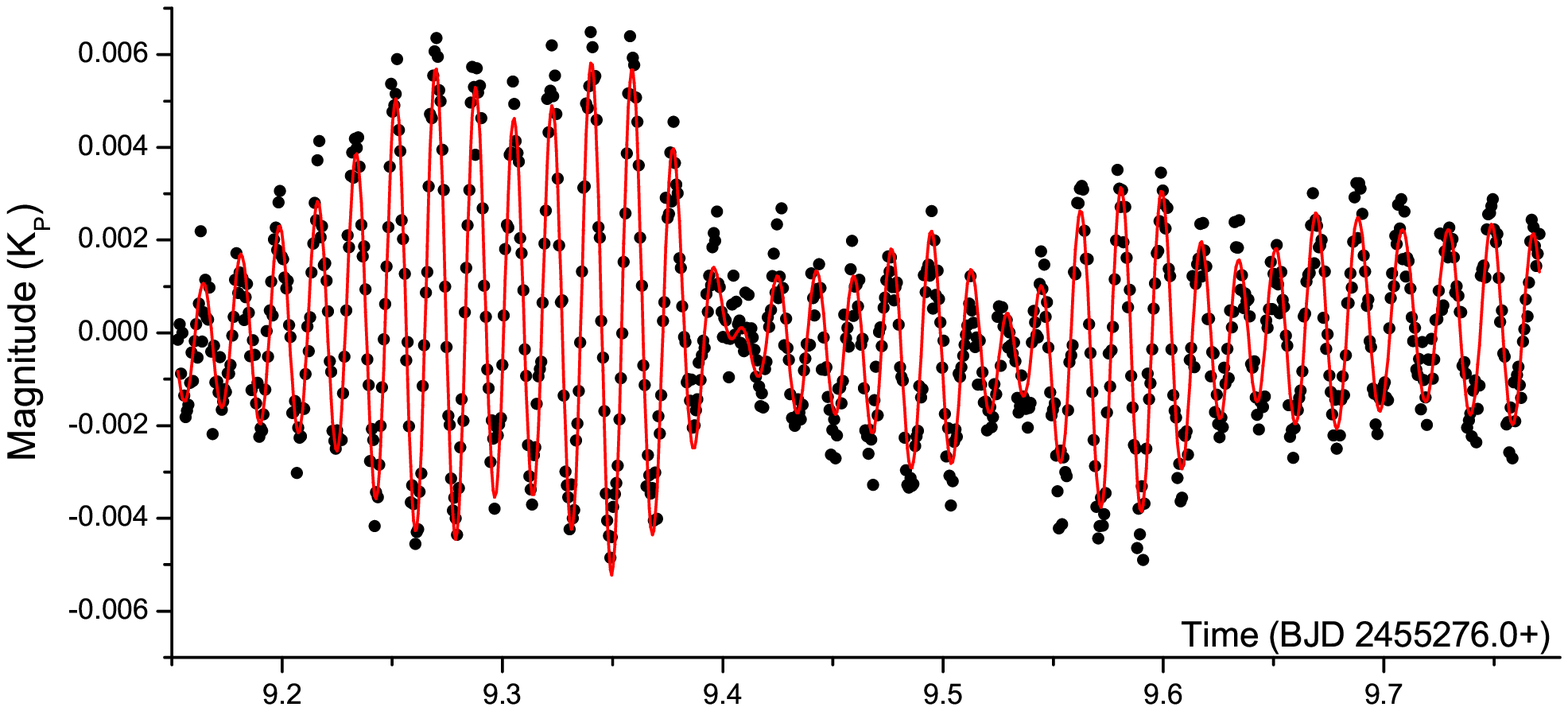}\\
\caption{Fourier fit (solid line) on various data points of Q5 for KIC~8553788.}
\label{fig:FF}
\includegraphics[width=8.5cm]{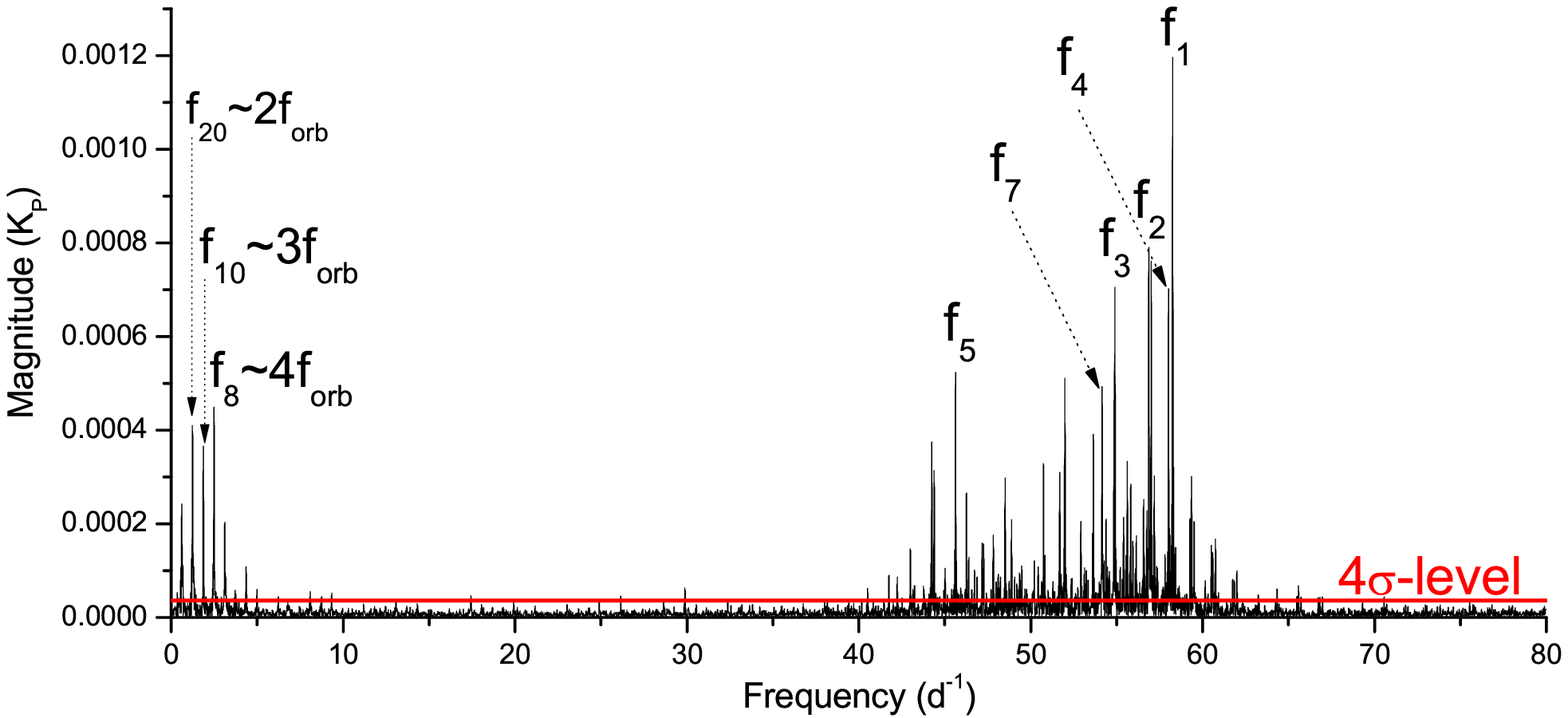}
\caption{Periodogram for KIC~8553788 for the Q5 data. The independent frequencies, the strong frequencies, that are connected to the $P_{\rm orb}$, and the 4-$\sigma$ significance level are also indicated. }
\label{fig:FS}
\end{figure}

Six independent frequencies were found in the range 45-59~cycle~d$^{-1}$, while another 83 were also detected and identified as combination or harmonics of others. The independent frequencies are listed in Table~\ref{tab:IndF}, while the others in Table~\ref{tab:DepFreq} in Appendix~\ref{sec:App1}. Both tables include the frequency values $f_{\rm i}$ (where $i$ is an increasing number), the semi-amplitudes $A$, the phases $\Phi$, and S/N. Additionally, the last two columns of Table~\ref{tab:IndF} include the pulsation constants ($Q$) and the $P_{\rm pul}/P_{\rm orb}$ ratio of each frequency, while the last column of Table~\ref{tab:DepFreq} contains the frequency combination. In Fig.~\ref{fig:FF} is presented the Fourier Fit on individual data points for two different days of observations, in Fig.~\ref{fig:FS} is illustrated the periodogram of the frequency search, while the bottom section of Fig.~\ref{fig:LCs1} shows the residuals after subtracting the Fourier model.

Although the system lacks of total eclipses, that would directly lead us to understand which of its components presents the pulsational activity (i.e. during the total eclipse of the pulsator the light variations due to the oscillations disappear), the most probable candidate for exhibiting those pulsations is the primary component. In particular, the evolutionary properties of this star, derived from the spectroscopic and LC analyses (Sects.~\ref{sec:sp} and \ref{sec:LCmdl}), and the frequencies ranges, where the independent frequencies exist, are consistent with the primary as a pulsating star of $\delta$~Sct type (see Section~\ref{sec:intro}).

\begin{table}
\centering
\caption{Independent oscillation frequencies for KIC~8553788.}
\label{tab:IndF}
\scalebox{0.85}{
\begin{tabular}{l cc cc cc}
\hline													
$i$	&	  $f_{\rm i}$	&	$A$	    &	  $\Phi$	&	S/N	  & 	$Q$	& $P_{\rm pul}/P_{\rm orb}$$^{\rm a}$	\\
	&(cycle~d$^{-1}$)	&	(mmag)	&	$(\degr)$	&		  & 	(d)  	&		\\
\hline													
1	&	58.2607(1)	&	1.196(4)	&	46.7(2)	&	130.3	&	0.008(1)	&	0.0107	\\
2	&	56.8702(1)	&	0.786(4)	&	14.0(3)	&	85.7	&	0.008(1)	&	0.0109	\\
3	&	54.8990(1)	&	0.707(4)	&	132.2(3)	&	77.0	&	0.008(1)	&	0.0113	\\
4	&	58.0183(1)	&	0.693(4)	&	112.7(4)	&	75.5	&	0.008(1)	&	0.0107	\\
5	&	45.6281(1)	&	0.523(4)	&	72.0(5)	&	56.9	&	0.010(2)	&	0.0136	\\
7	&	54.1584(2)	&	0.496(4)	&	83.6(5)	&	54.1	&	0.009(1)	&	0.0115	\\
\hline																			
\end{tabular}}
\newline
\textbf{Notes.} The errors are given in parentheses alongside values and correspond to the last digit(s). $^{(\rm a)}$ Error values are of $10^{-9}$ order of magnitude.
\end{table}

As shown in Fig.~\ref{fig:FS}, there are clearly two groups of frequencies; one is located in the range 0-5~cycle~d$^{-1}$ and the other in the range 42-61~cycle~d$^{-1}$. In total, 19 frequencies were detected in the fist regime, 68 in the second, and two more between 26-30~cycle~d$^{-1}$ (i.e. $f_{62}$ and $f_{74}$). The fact that a relatively large amount of frequencies were detected in the slow oscillations regime (0-5~cycle~d$^{-1}$) apparently arises the scenario for a $\delta$~Sct-$\gamma$~Dor hybrid behavior of the pulsator. However, none of these frequencies was identified as independent and their majority are connected to the orbital frequency of the system ($f_{\rm orb}$=0.62260~cycle~d$^{-1}$). Therefore, according to the criteria of \citet{UYT11} for characterizing possible hybrid pulsators and the present results, it can be concluded that the primary of this system is a pure $\delta$~Sct pulsating star.

The characterization of the oscillation modes (i.e. radial, non-radial) of the independent frequencies was based on two methods. The first concerns the comparison of the pulsation constants of the independent frequencies (see Table~\ref{tab:IndF}) with those predicted by the theoretical models of \citet{FIT81} for $M=1.5~M_{\sun}$. The second one compares the ratios $P_{\rm pul}/P_{\rm orb}$ for each frequency with those of other similar systems \citep{ZHA13}. For the calculations of $Q$ the relation of \citet{BRE00} was used:
\begin{equation}
\log Q = -\log f + 0.5 \log g + 0.1M_{\rm bol} + \log T_{\rm eff} - 6.456
\end{equation}
where $f$ is the pulsation frequency, and $\log g$, $M_{\rm bol}$, and $T_{\rm eff}$ denote the standard quantities (see Table~\ref{tab:LCmdlAbs}). All frequencies, except for $f_{5}$, were identified as non radial $p$-modes of $l=3$ degrees ($p7l3$), while $f_{5}$ as the seventh harmonic of the radial fundamental ($R7H$) mode. It should be noted that the ratio $f_{5}/f_{1}\sim0.78$ has the closest value to 0.77, that was defined by \citet{STE79} as the typical ratio for the radial fundamental to the first overtone mode. For the second method, regarding the empirical limit of \citet{ZHA13} (i.e. frequencies with $P_{\rm pul}/P_{\rm orb}<0.07$ potentially belong to the $p$-mode region), it was found that the independent frequencies are below that threshold, thus, may be simply $p$-modes.

According to the results of the frequency analysis (i.e. frequency values, frequency ranges) it can be plausibly concluded that the pulsating component of the system is a multiperiodic $\delta$~Sct star, that oscillates mainly in five pressure and probably in one radial modes accompanied with slower oscillations, that are related to the presence of the companion star.

\section{Evolution}
\label{sec:Evol}

This section includes all the possible scenarios for the physical and orbital evolution of the system under study.

\begin{figure}
\includegraphics[width=\columnwidth]{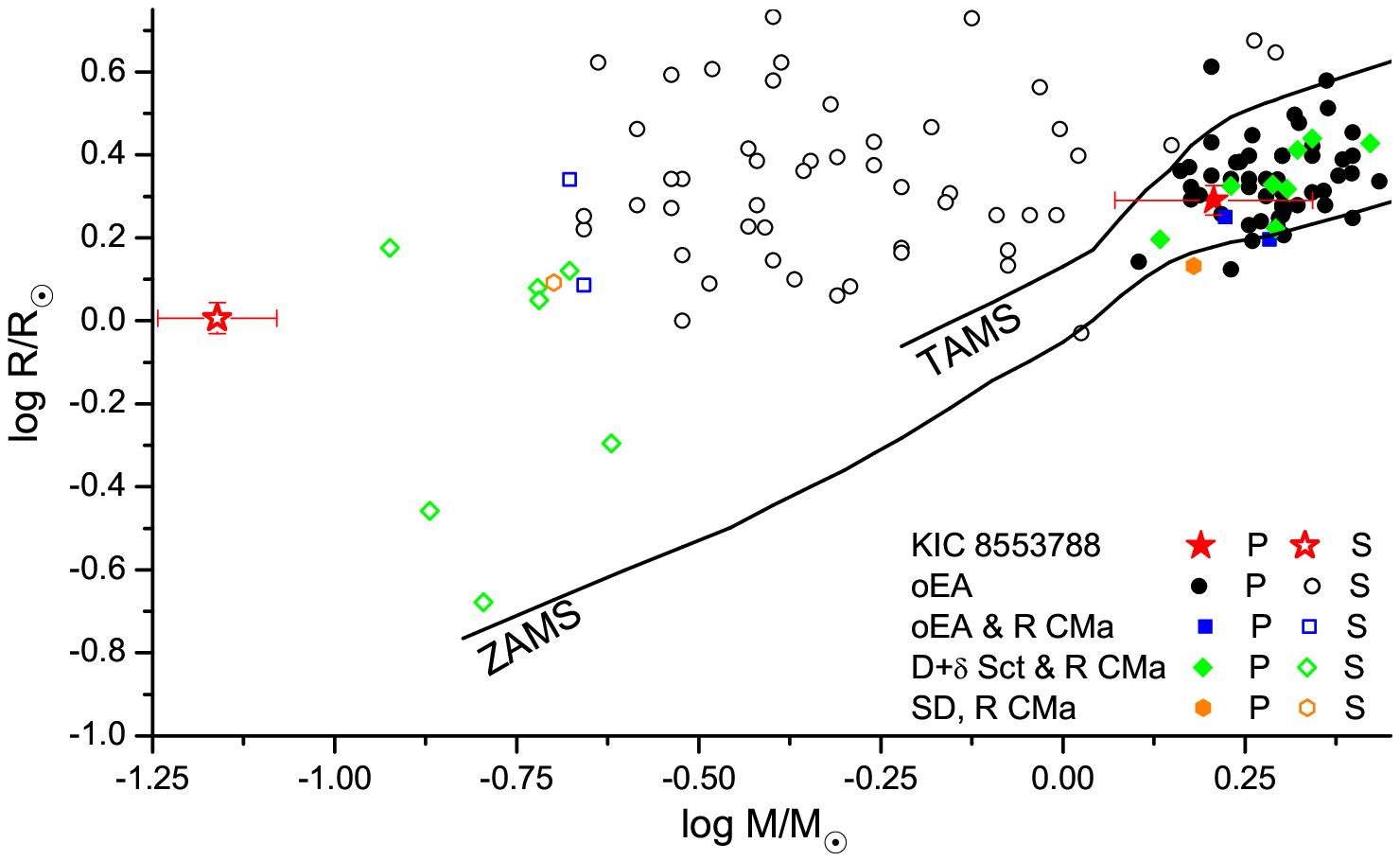}
\caption{Location of the components of KIC~8553788 (star symbols) within the Mass-Radius diagram among others of oEA systems (dot symbols) and others that belong to both oEA and R~CMa systems (square symbols). The diamond symbols denote the detached systems with a $\delta$~Scuti primary component that are also members of the R~CMa group, while the hexagon symbol the only semi-detached R~CMa system that does not exhibit pulsations. Filled symbols represent the primary, while the open ones the secondary components. Theoretical ZAMS and TAMS lines were taken from \citet{GIR00}.}
\label{fig:MR}
\includegraphics[width=\columnwidth]{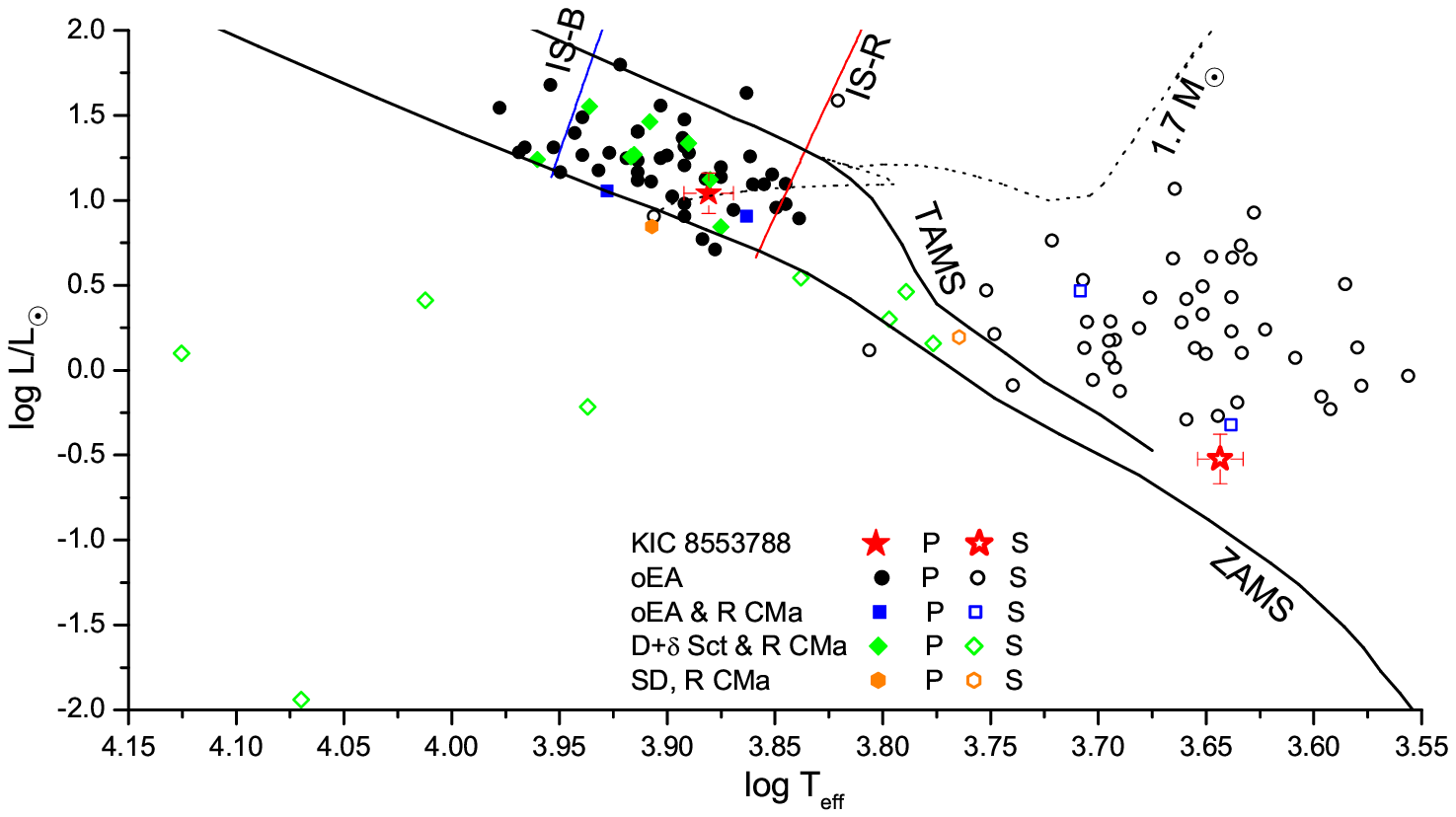}
\caption{Location of the components of KIC~8553788 within the HR diagram. Dashed line denotes the best fit theoretical model for the primary component and the coloured lines the Blue (B) and the Red (R) edges of the instability strip (IS). Symbols and rest lines represent the same as in Fig.~\ref{fig:MR}.}
\label{fig:HR}
\end{figure}

Figs.~\ref{fig:MR}-\ref{fig:HR} show the locations of both components of KIC~8553788 within the mass-radius ($M-R$) and Hertzsprung-Russell ($HR$) evolutionary diagrams, respectively, along with stars that belong to oEA and R~CMa systems. Zero and Terminal Age (ZAMS and TAMS, respectively) main-sequence lines (for solar metallicity composition) for these diagrams were taken from \citet{GIR00}, the positions of the stars of oEA systems from \citet{LIAN17}, and those of R~CMa systems from the literature (see Section~\ref{sec:intro} for individual system). In both diagrams is shown that the pulsator of the studied system lies well inside the main-sequence boundaries. The best theoretical evolutionary model \citep{GIR00} that fits better its position in the $HR$ diagram is that of a star with mass $M=1.7~M_{\sun}$, which is very close to the value derived in the present study (i.e. $M_{1}=1.6\pm0.4~M_{\sun}$) and it shows that this component is slightly more luminous for its mass. On the other hand, the secondary component was found oversized and overluminous for its mass  ($M_{2}=0.07\pm0.01~M_{\sun}$). It is located on the left side in the $M-R$ diagram far away from the other secondary components of oEA and R~CMa systems, and on the lower right part in the $HR$ diagram where it can be identified as the second less luminous component of all systems and the less luminous among the secondaries that are not yet in the WD stage.

The most possible explanation scenario for the system regarding its $P_{\rm orb}$, $q$, and the properties of the secondary component is the case-B of binary evolution on a nuclear time scale as given by \citet{LOO92}. According to this scenario, a mass ratio inverse took place in the system in the past. In particular, the present secondary component was initially the primary one (more massive) and its critical Roche lobe radius became larger than its stellar radius at the end of the core hydrogen burning but smaller than that at the beginning of He fusion. Therefore, the mass transfer began when the outer layers of the star were expanding, its luminosity decreased, reached a minimum and started to rise again. The core of the star may have became degenerated as the hydrogen shell decreased but given that this process was slow (i.e. slow decrease of core radius), probably there was no rapid expansion of the envelope, thus, no rapid mass transfer.

However, this very low $q$ value arises also the hypothesis of non-conservative mass transfer accompanied by mass loss due to magnetic braking from the system. It seems that this scenario can be potentially supported by the current finding of magnetic activity in the secondary component. Unfortunately, the ETV diagram of the system covers only approximately a decade of its life, therefore no sign of mass transfer/loss can be verified yet via steady increase/decrease of the $P_{\rm orb}$, but only cyclic changes that are discussed below. However, a further argue on the absence of secular changes of the $P_{\rm orb}$ in the ETV can be considered useful. For this the formula for secular period changes ($\Delta P_{\rm orb}/P_{\rm orb}$) suggested by \citet{ERD05}:
\begin{equation}
\frac{\Delta P_{\rm orb}}{P_{\rm orb}}=3k^2 (\frac{r_{\rm A}}{a})^2 \frac{m_1+m_2}{m_1 m_2} \dot{m}_{\rm loss} + 3 \frac{m_2-m_1}{m_1 m_2} \dot{m}_{\rm trans}
\end{equation}
where $k$ is the gyration constant of the mass donor, $a$ the semi-major axis of the EB, $\dot{m}_{\rm trans}$ the mass transfer rate from the less to the more massive component, and $\dot{m}_{\rm loss}$ the mass loss rate from the system up to the Alfv\'{e}n radius $r_{\rm A}$ will be used. Using the parameters of Table~\ref{tab:LCmdlAbs}, assuming typical values for the $k$ and $r_{\rm A}$ parameters \citep[cf. $k=0.32$, $r_{\rm A}=10a$,][]{ERD05}, and adopting a typical $\dot{m}_{\rm trans}$ for Algols (i.e. $10^{-8}~M_{\sun}$~yr$^{-1}$), that will lead to very low $\Delta P_{\rm orb}$ (i.e. $\sim5\times10^{-10}$~d~yr$^{-1}$) to be detected in a few years, a mass loss rate of $\sim10^{-9}~M_{\sun}$~yr$^{-1}$ is derived. Therefore, according to these assumptions, the hypothesis of non-conservative mass transfer may be possible but not yet detectable.

\citet{ZAS15} and \citet{BOR16} published the ETV analysis of the system resulting in the potential presence of a third body in a highly eccentric orbit. The mass functions for the tertiary ($f(m_{3})$) derived from these studies were 0.043~$M_{\sun}$ and 0.07~$M_{\sun}$, respectively. Moreover, \citet{ZAS15} assuming equal mass values for the components and a detached configuration for the system found a third light contribution of $\sim$6.6\%. \citet{BOR16}, based on an assumption of a total mass of 2~$M_{\sun}$ for the components of the EB, suggested a minimum mass of $m_{3, \rm min}\sim0.81$~$M_{\sun}$. In order to calculate the updated minimum mass of the third body, the derived parameters of the present study (Table~\ref{tab:LCmdlAbs}), and the third body's mass function formula of \citet{MAY90} were used. Assuming a coplanar orbit of the tertiary component, minimum masses of $m_{3, \rm min}$=0.61~$M_{\sun}$ and 0.74~$M_{\sun}$ were derived for each $f(m_{3})$ of the aforementioned studies, respectively. Thus, following the method of \citet{LIA11} (i.e. assume the main-sequence nature of the third component) and using the \textsc{InPeVEB} software \citep{LIA15}, it was found that for the first case the third light contribution should be $\sim1.5\%$, while for the second $\sim3\%$. These values can be considered, on one hand, as relatively low for the third component to be detected photometrically, but, on the other hand, not impossible. Therefore, the absence of a third light from the present photometric solution enables a few scenarios regarding the nature of the additional member. This component could be either an evolved star (e.g. WD), that is less luminous, or it could be a low mass binary system with late type components that emit an undetectable amount of light. However, since the mass of the tertiary can be increased significantly by rejecting the coplanar orbit assumption, it was found meaningless to speculate further on its nature.

Since it was found that there is magnetic activity in the system (see Section~\ref{sec:LCmdl}), the Applegate's mechanism \citep[i.e. quadrupole moment variation $\Delta Q$,][]{APP92} was also tested as an alternative explanation for the cyclic orbital period changes. For this, the ETV amplitudes and the periods of the cyclic modulation given in the works of \citet{ZAS15} and \citet{BOR16}, and the absolute parameters of the secondary component (Table~\ref{tab:LCmdlAbs}) were used in the formula of \citet{LAN02}, which is applied in the \textsc{InPeVEB} software \citep{LIA15}. The results showed that $\Delta Q$ is of the order of $10^{49}$~gr~cm$^2$, that is too low \citep[criterion established by][]{LAN02} to modulate the $P_{\rm orb}$ of the EB.

Concluding, a third component in a wide eccentric orbit is possible to exist, but so far neither its presence can be verified with direct measurements (photometry, spectroscopy) nor its nature. However, the low $q$, $P_{\rm orb}$ values of the EB can be somehow connected to this unseen member. According to \citet{BUD11}, conservative mass transfer scenarios for the R~CMa systems do not reflect their past and a non-conservative angular momentum redistribution between the tertiary component and the EB might be the explanation for the combination of low $q$, $P_{\rm orb}$.

\section{Discussion and Conclusions}
\label{sec:Dis}

This paper focused on the detailed analyses of KIC~8553788, an EB that was observed by the $Kepler$ mission. The SC data of $Kepler$ along with ground-based spectroscopic observations, which allowed the estimation of the spectral type of the brighter star of the system, and the literature RV curve of the primary component provided the means for the most accurate photometric modelling that can be performed to date. Consequently, the absolute parameters of both components were derived revealing a very low mass ratio for the system ($q=0.043$). Moreover, it was found that the primary component is a $\delta$~Scuti type star that pulsates in six independent frequencies and in other 83 depended ones or harmonics of others. The secondary component was found to have an extremely low mass value ($M_{2}=0.07~M_{\sun}$), while migrating photospheric spots on its surface were also detected.

Comparison between the results from the present LC analysis and those of \citet{ZAS15} shows large discrepancies due to the different approach followed in each case. In particular, the temperature ratio ($T_{2}/T_{1}$) and the inclination were found as $\sim$0.58 and $\sim75\degr$ in the present study, respectively, while \citet{ZAS15} found $\sim$0.64 and $\sim70\degr$, respectively. In addition, \citet{ZAS15} found a third light contribution of $\sim$6.6\%, something that was not confirmed by the present results. The main reason for these differences concerns the inclusion of spectroscopic data in the present analysis. \citet{ZAS15} assumed a priori a detached system with $q=1$, while the present results are strongly based on the RV curve \citep{MAT17} and the spectroscopically determined temperature (see Section~\ref{sec:sp}) of the primary component and on the ``$q$-search'' method.

The system was found to be in conventional semi-detached status, thus, according to the definition of oEA stars \citep{MKR02}, it can be considered as a member of this group. Moreover, its classical Algol type status, its relatively short $P_{\rm orb}$ and its low $q$ are characteristics that lie well inside the regime of properties of the group of R~CMa type systems as defined by \citet{BUD11}. The rarity of this system concerns the fact that it is a member of two different groups at the same time, with only two other systems, namely R~CMa and AS~Eri, to have similar characteristics (see Fig.~\ref{fig:groups}).

\begin{figure}
\includegraphics[width=\columnwidth]{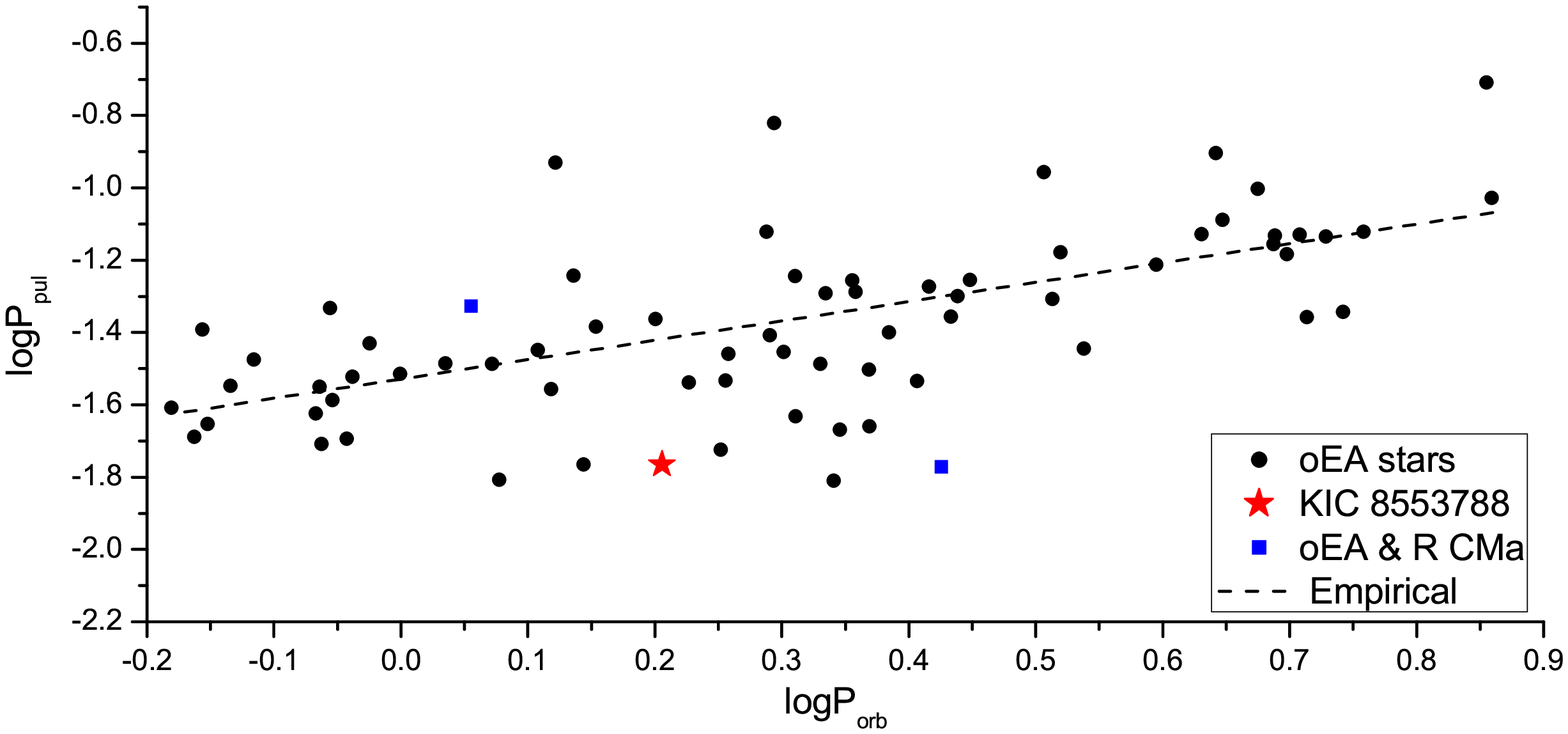}
\caption{Location of the pulsating component of KIC~8553788 within the $P_{\rm orb}-P_{\rm pul}$ diagram among others of oEA systems and others that belong to both oEA and R~CMa systems. The star, dot, and square symbols have the same meaning as in Fig.~\ref{fig:MR}. Dashed line denotes the empirical fit of \citet{LIAN17}.}
\label{fig:PP}
\includegraphics[width=\columnwidth]{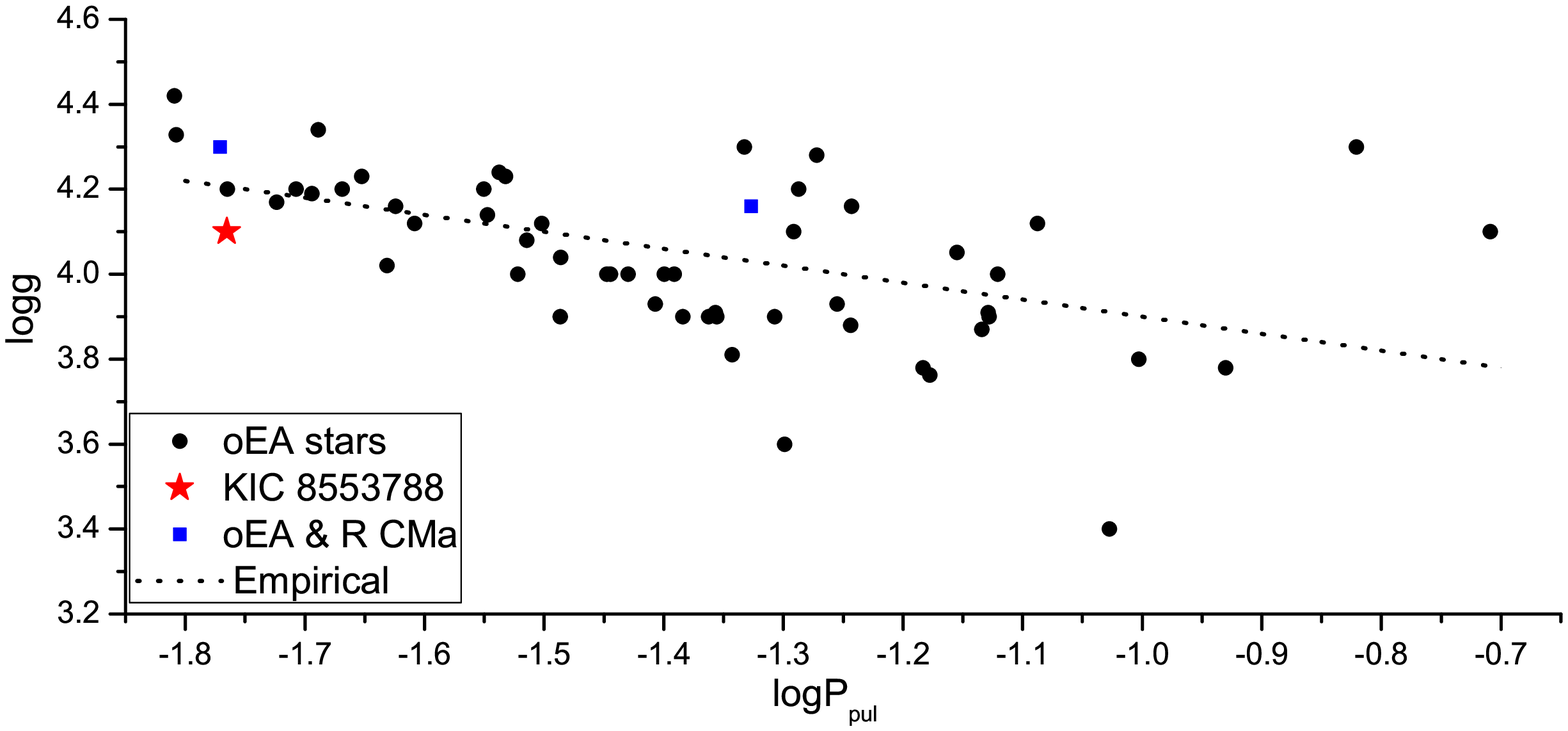}
\caption{Location of the pulsating component of KIC~8553788 within the $\log g-P_{\rm pul}$ diagram among other stars-members of systems of similar types. Symbols and line represent the same as in Fig.~\ref{fig:PP}.}
\label{fig:gP}
\end{figure}

The primary component of the system was found to be the 4th fastest $\delta$~Scuti star ($f_{1}=58.26$~cycle~d$^{-1}$) among the oEA systems that has been found to date. According to the catalogue of \citet{LIAN17}, there are only another four oEA systems with components that pulsate between 58-65~cycle~d$^{-1}$, namely KIC~11175495 \citep[64.44~cycle~d$^{-1}$,][]{LIA17}, RZ~Cas \citep[64.20~cycle~d$^{-1}$,][]{SOY07}, AS~Eri \citep[59.03~cycle~d$^{-1}$,][]{MKR04, IBA06}, and AU~Lac \citep[58.22~cycle~d$^{-1}$,][]{LIA12}, while below that range there is a gap between 53-58~cycle~d$^{-1}$, where no systems have been found. In Fig.~\ref{fig:PP} is shown the location of the pulsator of KIC~8553788 within the $P_{\rm orb}-P_{\rm pul}$ correlation diagram for oEA stars along with the empirical fit of \citet{LIAN17}. The system is one of the most deviating cases from the fit, because of its very fast pulsating frequency that does not follow the trend between orbital and pulsation periods. One possible explanation for this system, but also for the other deviating ones (i.e. very fast pulsators), might be the mass accretion from the secondary components. However, this cannot be verified yet, since the ETV diagrams of the aforementioned systems (c.f. \citet[][for RZ~Cas, AS~Eri, and AU~Lac]{KRE01}, \citet[][for KIC~11175495]{GIE15}) do not show any significant parabolic trend that could be attributed to mass transfer. In Fig.~\ref{fig:gP} is given the distribution of the pulsators of oEA systems in the $\log g - \log P_{\rm pul}$ diagram. The oscillating member of KIC~8553788, although is very close to the fit of \citet{LIAN17} \citep[updated by][]{LIA17}, is the only one below that, in contrast with the other four similar stars (upper left part of Fig.~\ref{fig:gP}). If confirmed by spectroscopic measurements of the secondary component, this system would have the smallest mass ratio among these four EBs and although its primary has the smallest mass among the other pulsators, it would be the more evolved. This fact probably shows that the mass transfer in KIC~8553788 has greater rate or began earlier in comparison with the other systems and has led the primary to evolve faster.

Possible evolutionary scenarios with respect to the present status of the system were discussed. Although there is lack of evidence in its ETV diagram of a secular modulating mechanism of the $P_{\rm orb}$, it seems that non-conservative mass transfer, mass loss from the system, and/or angular momentum redistribution from the EB to the tertiary component are the mechanisms that drove the system to have this extremely low mass ratio value.

Comparison of the mass ratio of  KIC~8553788 was made with the semi-detached systems included in the catalogues of \citet{BUD04}, \citet{IBA06}, and \citet{SOY06b}. Results showed that, indeed, this EB has the lowest value among them, therefore it can be fairly considered as a potentially (i.e. RV data of the secondary component are needed to confirm it) extreme case \citep[cf.][]{GUL16, DUC11, BOF10}.

Undoubtedly, KIC~8553788 is a very rare case of binary systems, in terms of evolutionary and physical properties. In can be considered as a valuable case for the binary evolution modelling theorists, especially for its very low mass ratio. Long-term monitoring (e.g. decades) of the system for eclipse timings acquisitions probably will shed light to its orbital evolution and help us to check further the current evolutionary scenarios. In general, both the groups of oEA and R~CMa systems are still small, therefore any enrichment can be considered as very valuable and scientifically extremely important.

\begin{acknowledgements}
The author acknowledges financial support by the European Space Agency (ESA) under the Near Earth object Lunar Impacts and Optical TrAnsients (NELIOTA) programme, contract no. 4000112943. The author also wishes to thank Dr. Panayotis Boumis for his suggestions, Mrs Maria Pizga and Mrs Joanne Barton for proofreading the text, and the anonymous reviewer for the valuable comments. The ``Aristarchos'' telescope is operated on Helmos Observatory by the Institute for Astronomy, Astrophysics, Space Applications and Remote Sensing of the National Observatory of Athens. This research has made use of NASA's Astrophysics Data System Bibliographic Services, the SIMBAD, the Mikulski Archive for Space Telescopes (MAST), and the $Kepler$ Eclipsing Binary Catalog databases.
\end{acknowledgements}

%
%


\begin{appendix}

\section{List of combined frequencies}
\label{sec:App1}

This Appendix includes in Table~\ref{tab:DepFreq} the depended frequency values $f_{\rm i}$ (where $i$ is an increasing number), semi-amplitudes $A$, phases $\Phi$, S/N and their combinations. Details for the method followed for their derivation can be found in Section~\ref{sec:Fmdl}.

\begin{table*}
\centering
\caption{Combined frequencies of KIC~8553788. The errors are given in parentheses alongside values and correspond to the last digit(s).}
\label{tab:DepFreq}
\begin{tabular}{cc cc cc| cc cc cc}
\hline																							
$i$	&	  $f_{\rm i}$	&	$A$	&	  $\Phi$	&	S/N	&	Combination	&	$i$	&	  $f_{\rm i}$	&	$A$	&	  $\Phi$	&	S/N	&	Combination	\\
	&	     (cycle~d$^{-1}$)	&	(mmag)	&	$(\degr)$	&		&		&		&	     (cycle~d$^{-1}$)	&	(mmag)	&	$(\degr)$	&		&		\\
\hline																							
6	&	51.9905(1)	&	0.512(4)	&	157(1)	&	55.8	&	$f_{1}+2f_{3}-2f_{4}$	&	49	&	47.5073(10)	&	0.078(4)	&	280(3)	&	8.5	&	$f_{10}+f_{5}$	\\
8	&	2.4898(2)	&	0.451(4)	&	67(1)	&	49.2	&	$4f_{\rm orb}$	&	50	&	51.6702(10)	&	0.078(4)	&	325(3)	&	8.5	&	$\sim f_{13}$	\\
9	&	57.0161(2)	&	0.408(4)	&	328(1)	&	44.5	&	$f_{6}+2f_{8}$	&	51	&	60.1577(10)	&	0.076(4)	&	28(3)	&	8.3	&	$f_{10}+f_{1}$	\\
10	&	1.8665(2)	&	0.364(4)	&	304(1)	&	39.7	&	$3f_{\rm orb}$	&	52	&	43.2477(10)	&	0.075(4)	&	90(3)	&	8.1	&	$f_{10}+f_{9}-f_{4}$	\\
11	&	54.8451(2)	&	0.361(4)	&	308(1)	&	39.3	&	$f_{6}+f_{9}-f_{7}$	&	53	&	51.2912(10)	&	0.074(4)	&	45(3)	&	8.0	&	$f_{6}-f_{29}$	\\
12	&	44.2468(2)	&	0.355(4)	&	48(1)	&	38.6	&	$f_{2}+f_{5}-f_{1}$	&	54	&	56.1534(10)	&	0.072(4)	&	69(3)	&	7.8	&	$f_{4}-f_{10}$	\\
13	&	51.6924(2)	&	0.349(4)	&	101(1)	&	38.0	&	$f_{7}-f_{8}$	&	55	&	58.4465(11)	&	0.071(4)	&	204(3)	&	7.7	&	$f_{10}+f_{19}$	\\
14	&	58.1135(2)	&	0.322(4)	&	283(1)	&	35.1	&	$f_{1}+f_{2}-f_{9}$	&	56	&	1.3131(11)	&	0.069(4)	&	23(4)	&	7.5	&	$2f_{29}$	\\
15	&	48.5143(3)	&	0.297(4)	&	129(1)	&	32.4	&	$f_{3}+f_{5}-f_{6}$	&	57	&	53.9871(11)	&	0.069(4)	&	238(4)	&	7.5	&	$f_{18}-f_{10}$	\\
16	&	57.1937(3)	&	0.279(4)	&	266(1)	&	30.4	&	$f_{1}+f_{9}-f_{14}$	&	58	&	48.4239(11)	&	0.068(4)	&	107(4)	&	7.4	&	$f_{20}+f_{26}$	\\
17	&	46.2624(3)	&	0.258(4)	&	25(1)	&	28.1	&	$f_{12}+f_{2}-f_{3}$	&	59	&	4.3579(11)	&	0.066(4)	&	262(4)	&	7.1	&	$\sim 7f_{\rm orb}$	\\
18	&	55.8251(3)	&	0.258(4)	&	255(1)	&	28.1	&	$f_{16}+f_{2}-f_{1}$	&	60	&	47.271(12)	&	0.064(4)	&	10(4)	&	7.0	&	$f_{15}-f_{20}$	\\
19	&	56.5784(3)	&	0.242(4)	&	7(1)	&	26.3	&	$2f_{2}-f_{16}$	&	61	&	55.4049(12)	&	0.062(4)	&	125(4)	&	6.8	&	$f_{20}+f_{7}$	\\
20	&	1.2290(3)	&	0.235(4)	&	114(1)	&	25.6	&	$f_{1}-f_{9} \sim 2f_{\rm orb}$	&	62	&	29.8854(12)	&	0.061(4)	&	127(4)	&	6.7	&	$f_{48}+f_{5}-f_{4}$	\\
21	&	48.8902(4)	&	0.205(4)	&	223(1)	&	22.3	&	$f_{3}+f_{6}-f_{4}$	&	63	&	47.696(13)	&	0.059(4)	&	132(4)	&	6.4	&	$f_{33}-f_{8}$	\\
22	&	53.6557(4)	&	0.196(4)	&	50(1)	&	21.3	&	$f_{3}-f_{20}$	&	64	&	1.1989(14)	&	0.055(4)	&	78(4)	&	6.0	&	$\sim f_{20}$	\\
23	&	0.6058(4)	&	0.192(4)	&	241(1)	&	20.9	&	$f_{8}-f_{10} \sim f_{\rm orb}$	&	65	&	0.3505(14)	&	0.054(4)	&	176(5)	&	5.9	&	$f_{16}-f_{2}$	\\
24	&	47.8261(4)	&	0.178(4)	&	318(1)	&	19.4	&	$f_{5}+f_{7}-f_{6}$	&	66	&	0.5788(14)	&	0.053(4)	&	148(5)	&	5.7	&	$\sim f_{23}$	\\
25	&	50.7456(4)	&	0.175(4)	&	87(1)	&	19.1	&	$f_{6}-f_{20}$	&	67	&	49.7846(14)	&	0.052(4)	&	305(5)	&	5.7	&	$f_{13}-f_{10}$	\\
26	&	47.1791(5)	&	0.155(4)	&	197(2)	&	16.9	&	$f_{24}-f_{23}$	&	68	&	52.6471(15)	&	0.052(4)	&	137(5)	&	5.7	&	$f_{29}+f_{6}$	\\
27	&	44.3832(5)	&	0.147(4)	&	353(2)	&	16.0	&	$f_{5}-f_{20}$	&	69	&	44.9462(15)	&	0.052(4)	&	4(5)	&	5.6	&	$f_{5}-f_{29}$	\\
28	&	49.4881(6)	&	0.135(4)	&	29(2)	&	14.8	&	$f_{6}-f_{8}$	&	70	&	57.8249(15)	&	0.049(4)	&	55(5)	&	5.3	&	$f_{19}+f_{20}$	\\
29	&	0.6676(6)	&	0.132(4)	&	299(2)	&	14.3	&	$f_{11}-f_{7}$	&	71	&	54.1822(16)	&	0.048(4)	&	339(5)	&	5.3	&	$\sim f_{7}$	\\
30	&	58.2626(6)	&	0.125(4)	&	228(2)	&	13.6	&	$\sim f_{1}$	&	72	&	56.9003(16)	&	0.048(4)	&	122(5)	&	5.2	&	$\sim f_{2}$	\\
31	&	1.2592(6)	&	0.125(4)	&	178(2)	&	13.6	&	$\sim 2f_{\rm orb}$	&	73	&	51.7178(16)	&	0.047(4)	&	235(5)	&	5.1	&	$\sim f_{13}$	\\
32	&	45.4901(6)	&	0.122(4)	&	314(2)	&	13.3	&	$f_{12}+f_{20}$	&	74	&	26.1523(16)	&	0.047(4)	&	63(5)	&	5.1	&	$f_{62}-f_{41}$	\\
33	&	50.2160(7)	&	0.114(4)	&	75(2)	&	12.4	&	$f_{1}+f_{5}-f_{22}$	&	75	&	1.2830(16)	&	0.047(4)	&	320(5)	&	5.1	&	$\sim f_{31}$	\\
34	&	55.2193(7)	&	0.111(4)	&	211(2)	&	12.0	&	$f_{18}-f_{23}$	&	76	&	1.3638(16)	&	0.047(4)	&	86(5)	&	5.1	&	$2f_{29}$	\\
35	&	57.0716(7)	&	0.106(4)	&	133(2)	&	11.5	&	$f_{1}-f_{20}$	&	77	&	46.3956(17)	&	0.043(4)	&	267(6)	&	4.7	&	$f_{21}-f_{8}$	\\
36	&	56.8591(7)	&	0.108(4)	&	87(2)	&	11.8	&	$\sim f_{2}$	&	78	&	45.6329(18)	&	0.043(4)	&	233(6)	&	4.7	&	$\sim f_{5}$	\\
37	&	54.1346(7)	&	0.104(4)	&	133(2)	&	11.4	&	$\sim f_{7}$	&	79	&	56.4801(18)	&	0.043(4)	&	295(6)	&	4.7	&	$f_{2}-f_{65}$	\\
38	&	0.0222(8)	&	0.099(4)	&	231(2)	&	10.8	&	$f_{3}-f_{11}$	&	80	&	58.8430(18)	&	0.043(4)	&	44(6)	&	4.6	&	$f_{23}+f_{1}$	\\
39	&	47.6453(8)	&	0.098(4)	&	321(3)	&	10.6	&	$f_{28}-f_{10}$	&	81	&	1.8951(18)	&	0.042(4)	&	62(6)	&	4.6	&	$\sim f_{10}$	\\
40	&	3.1130(8)	&	0.096(4)	&	14(3)	&	10.5	&	$\sim 5f_{\rm orb}$	&	82	&	50.6029(18)	&	0.041(4)	&	337(6)	&	4.5	&	$f_{6}-f_{76}$	\\
41	&	3.7331(8)	&	0.097(4)	&	341(3)	&	10.5	&	$\sim 6f_{\rm orb}$	&	83	&	54.2186(19)	&	0.040(4)	&	338(6)	&	4.4	&	$\sim f_{71}$	\\
42	&	50.8344(8)	&	0.094(4)	&	293(3)	&	10.3	&	$f_{23}+f_{33}$	&	84	&	0.4123(19)	&	0.040(4)	&	148(6)	&	4.3	&	$f_{9}-f_{19}$	\\
43	&	54.8942(8)	&	0.094(4)	&	339(3)	&	10.2	&	$\sim f_{3}$	&	85	&	45.2824(19)	&	0.039(4)	&	291(6)	&	4.3	&	$f_{26}-f_{10}$	\\
44	&	60.6191(9)	&	0.088(4)	&	14(3)	&	9.6	&	$f_{14}+f_{8}$	&	86	&	53.5447(19)	&	0.039(4)	&	309(6)	&	4.2	&	$f_{10}+f_{13}$	\\
45	&	59.0745(9)	&	0.085(4)	&	80(3)	&	9.3	&	$f_{19}+f_{8}$	&	87	&	57.0050(20)	&	0.038(4)	&	357(6)	&	4.2	&	$\sim f_{9}$	\\
46	&	55.9488(9)	&	0.084(4)	&	176(3)	&	9.2	&	$f_{16}-f_{20}$	&	88	&	48.1972(20)	&	0.038(4)	&	32(7)	&	4.1	&	$f_{21}-f_{29}$	\\
47	&	0.5471(9)	&	0.080(4)	&	107(3)	&	8.7	&	$f_{7}-f_{22}$	&	89	&	52.1237(20)	&	0.037(4)	&	225(7)	&	4.1	&	$f_{10}+f_{33}$	\\
48	&	42.2423(9)	&	0.080(4)	&	21(3)	&	8.7	&	$f_{3}+f_{5}-f_{1}$	&		&		&		&		&		&		\\
\hline	
\end{tabular}
\end{table*}

\section{Spot migration}
\label{sec:App2}
This appendix includes information for the spots migration for the studied system during the Q5 and Q14 quarters of the $Kepler$ mission. Tables~\ref{tab:spots}~\ref{tab:spots2} include the co-latitude ($co-lat.$), longitude ($long.$), radius, and temperature factor ($Tf$) of each spot. The first timing given in the $Time$ columns is the starting time of the respective dataset plus the half of the $P_{\rm orb}$ of the EB. Each following timing has been calculated by simply adding the $P_{\rm orb}$ value to the previous one.

\begin{table}
\centering
\caption{Spot parameters for KIC~8553788 during the Q5.}
\label{tab:spots}
\begin{tabular}{c cc cc }
\hline			
Time	&	Co-lat	&	Long	&	Radius	&	Tf ($\frac{T_{\rm spot}}{T_{\rm eff}})$		\\
(BJD 2455276.0+)	&	($\degr$)	&	($\degr$)	&	($\degr$)	&				\\
\hline									
0.480	&	73.88	&	192.43	&	11.07	&	0.84	\\
2.087	&	73.88	&	184.68	&	11.07	&	0.85	\\
3.693	&	73.97	&	195.44	&	12.49	&	0.86	\\
5.299	&	73.97	&	189.64	&	12.49	&	0.85	\\
6.905	&	73.97	&	197.58	&	12.49	&	0.85	\\
8.511	&	73.97	&	196.79	&	12.49	&	0.86	\\
10.117	&	73.97	&	205.96	&	12.25	&	0.86	\\
11.724	&	73.34	&	202.79	&	12.89	&	0.87	\\
13.330	&	73.06	&	198.71	&	13.33	&	0.89	\\
14.936	&	70.46	&	203.67	&	13.97	&	0.91	\\
16.542	&	64.02	&	197.12	&	15.30	&	0.91	\\
18.148	&	64.02	&	198.20	&	15.27	&	0.90	\\
19.754	&	60.13	&	211.86	&	15.24	&	0.86	\\
21.361	&	64.35	&	222.46	&	13.87	&	0.84	\\
22.967	&	69.23	&	233.44	&	14.76	&	0.86	\\
24.573	&	68.99	&	241.58	&	15.18	&	0.89	\\
26.179	&	69.78	&	249.96	&	15.18	&	0.85	\\
27.785	&	70.06	&	250.03	&	14.71	&	0.86	\\
29.391	&	69.28	&	260.35	&	14.49	&	0.80	\\
\hline																												
\end{tabular}
\end{table}

\begin{table}
\centering
\caption{Spot parameters for KIC~8553788 during the Q14.}
\label{tab:spots2}
\begin{tabular}{c cc cc }
\hline			
Time	&	Co-lat	&	Long	&	Radius	&	Tf ($\frac{T_{\rm spot}}{T_{\rm eff}})$		\\
(BJD 2456107.0+)	&	($\degr$)	&	($\degr$)	&	($\degr$)	&				\\
\hline
\multicolumn{5}{c}{spot 1}\\
\hline									
0.131	&	57.20	&	125.97	&	26.10	&	0.94	\\
1.737	&	59.37	&	110.49	&	26.85	&	0.93	\\
3.343	&	59.52	&	105.59	&	26.88	&	0.92	\\
4.949	&	62.38	&	105.62	&	25.62	&	0.91	\\
6.555	&	57.70	&	96.97	&	24.46	&	0.92	\\
8.161	&	70.21	&	95.59	&	23.73	&	0.92	\\
9.768	&	64.19	&	91.71	&	23.74	&	0.91	\\
11.374	&	61.63	&	94.17	&	22.97	&	0.90	\\
12.980	&	58.53	&	88.72	&	22.51	&	0.88	\\
20.880	&	56.94	&	72.77	&	22.71	&	0.86	\\
22.486	&	55.50	&	64.39	&	24.82	&	0.91	\\
24.092	&	55.82	&	57.00	&	25.13	&	0.92	\\
25.699	&	55.57	&	61.82	&	24.88	&	0.92	\\
27.305	&	55.29	&	67.55	&	24.69	&	0.88	\\
\hline
\multicolumn{5}{c}{spot 2}\\
\hline									
0.131	&	93.02	&	205.51	&	18.92	&	0.73	\\
1.737	&	90.98	&	199.53	&	18.42	&	0.78	\\
3.343	&	90.98	&	197.64	&	18.37	&	0.80	\\
4.949	&	91.69	&	190.98	&	17.66	&	0.78	\\
6.555	&	91.53	&	188.33	&	17.82	&	0.75	\\
8.161	&	91.44	&	190.55	&	18.03	&	0.73	\\
9.768	&	89.61	&	187.09	&	17.22	&	0.71	\\
11.374	&	89.88	&	189.91	&	17.79	&	0.73	\\
12.980	&	90.47	&	185.48	&	19.45	&	0.76	\\
20.880	&	91.78	&	175.86	&	20.58	&	0.78	\\
22.486	&	91.74	&	181.21	&	20.54	&	0.75	\\
24.092	&	89.69	&	185.91	&	21.13	&	0.76	\\
25.699	&	89.44	&	184.75	&	20.88	&	0.79	\\
27.305	&	88.61	&	178.65	&	20.05	&	0.82	\\
\hline																																	
\end{tabular}
\end{table}

\begin{figure}
\begin{tabular}{cc}
\multicolumn{2}{c}{\includegraphics[width=7.5cm]{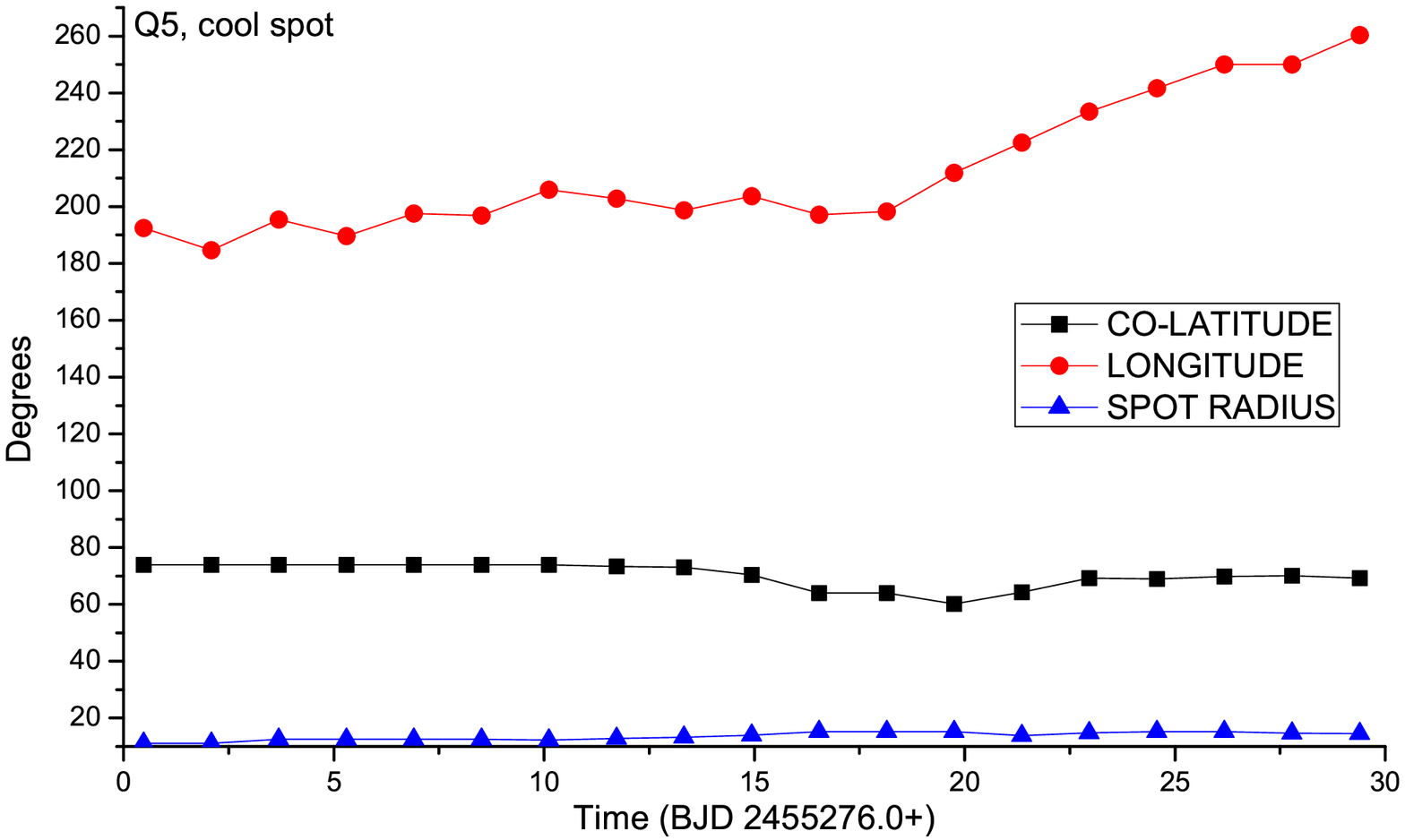}}\\
\includegraphics[width=3.7cm]{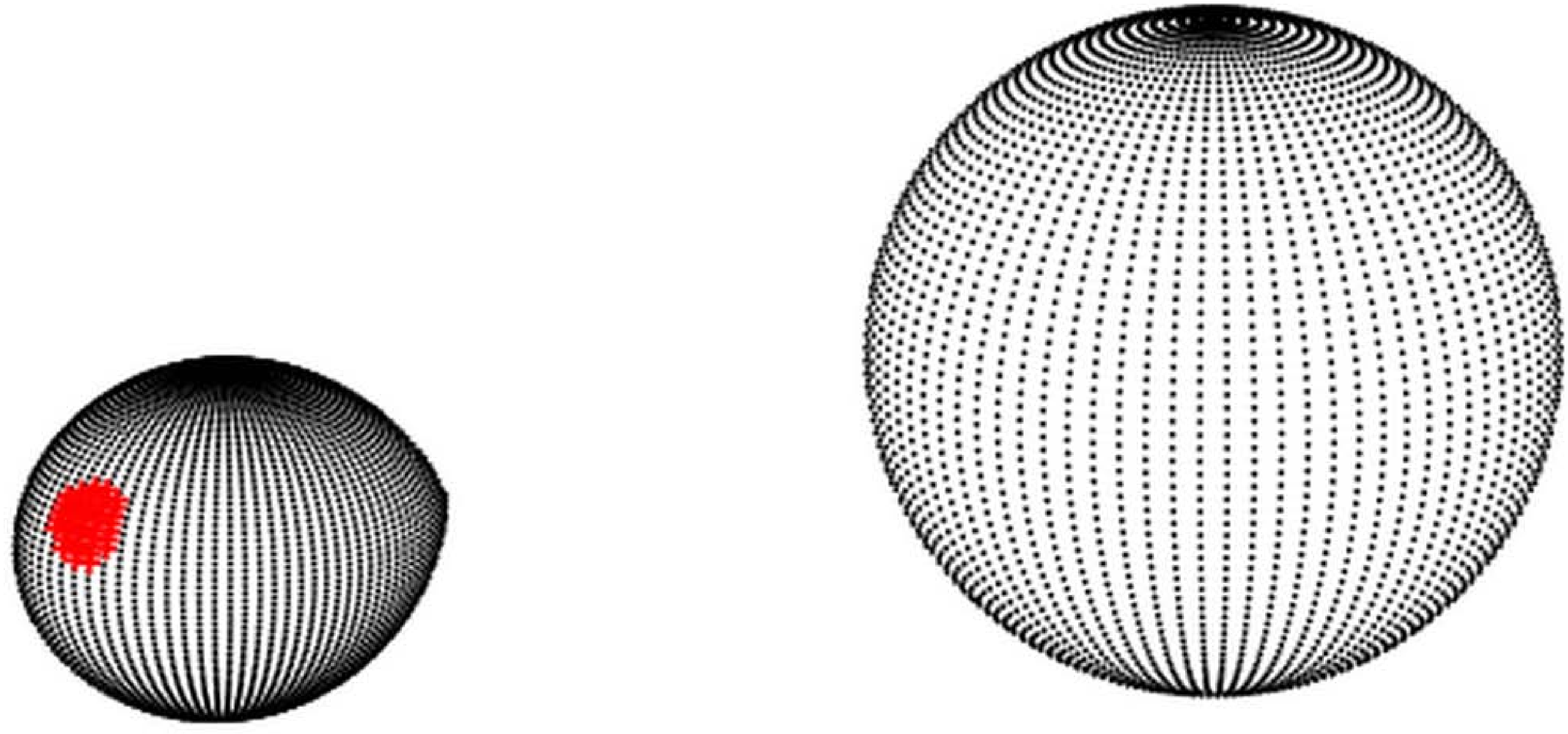}&\hspace{0.5cm}\includegraphics[width=3.7cm]{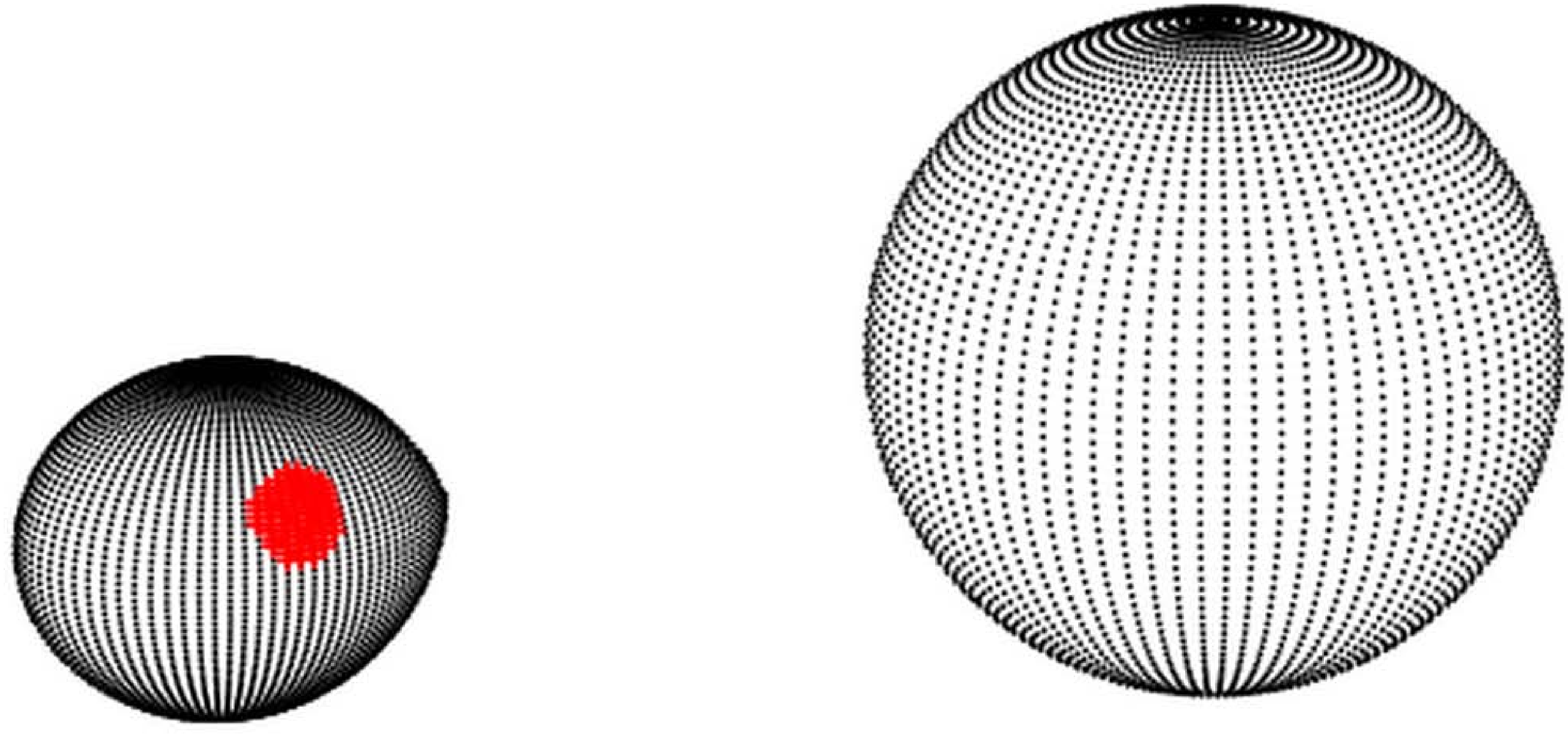}\\
\end{tabular}
\caption{Upper panel: Spot migration diagram for KIC~8553788 during Q5. Lower panels: The location of the spot (crosses) on the surface of the secondary component during the first day (left) and the last day (right), when the system is at orbital phase 0.85.}
\label{fig:spotmigr1}

\begin{tabular}{cc}
\multicolumn{2}{c}{\includegraphics[width=7.5cm]{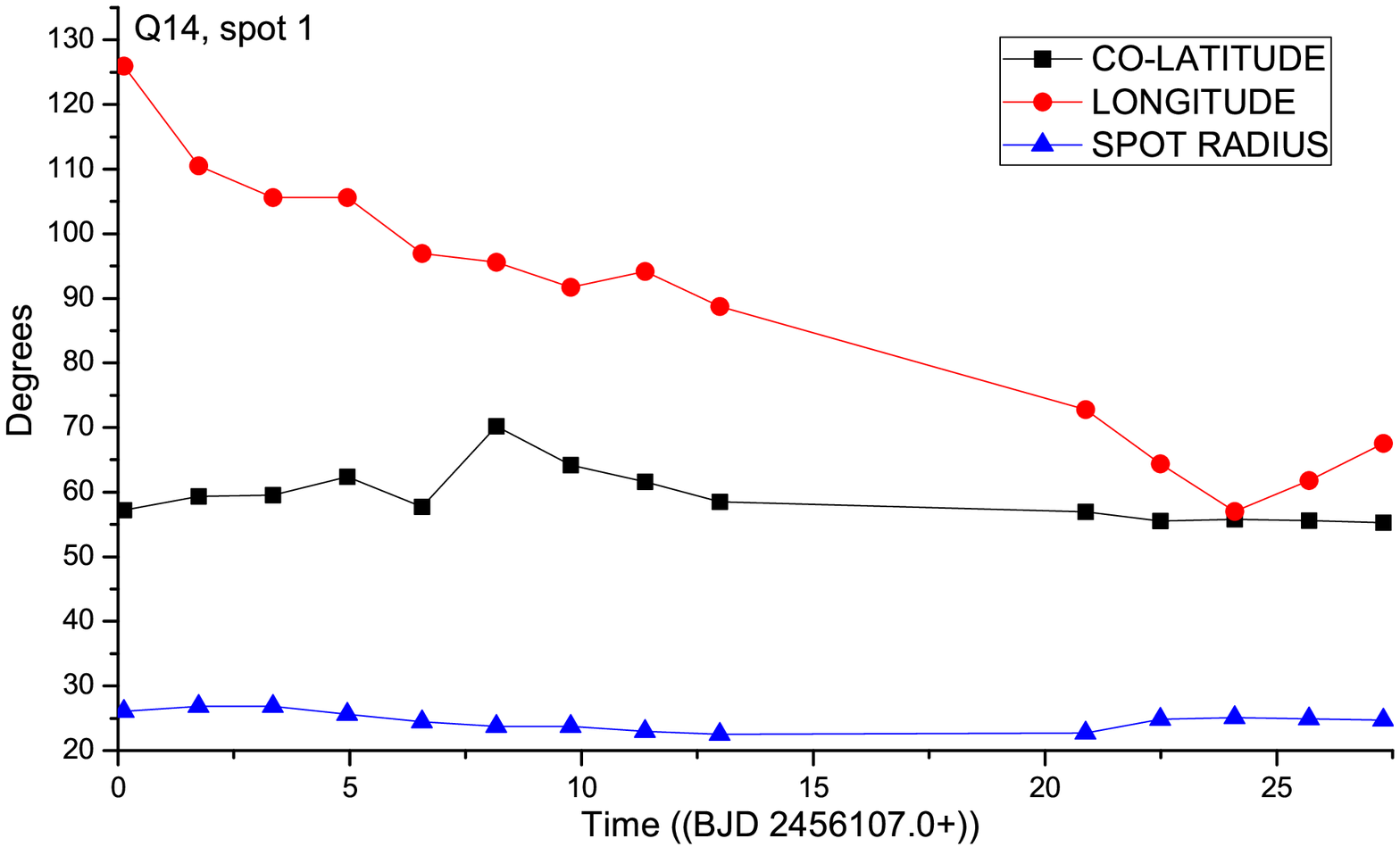}}\\
\includegraphics[width=3.7cm]{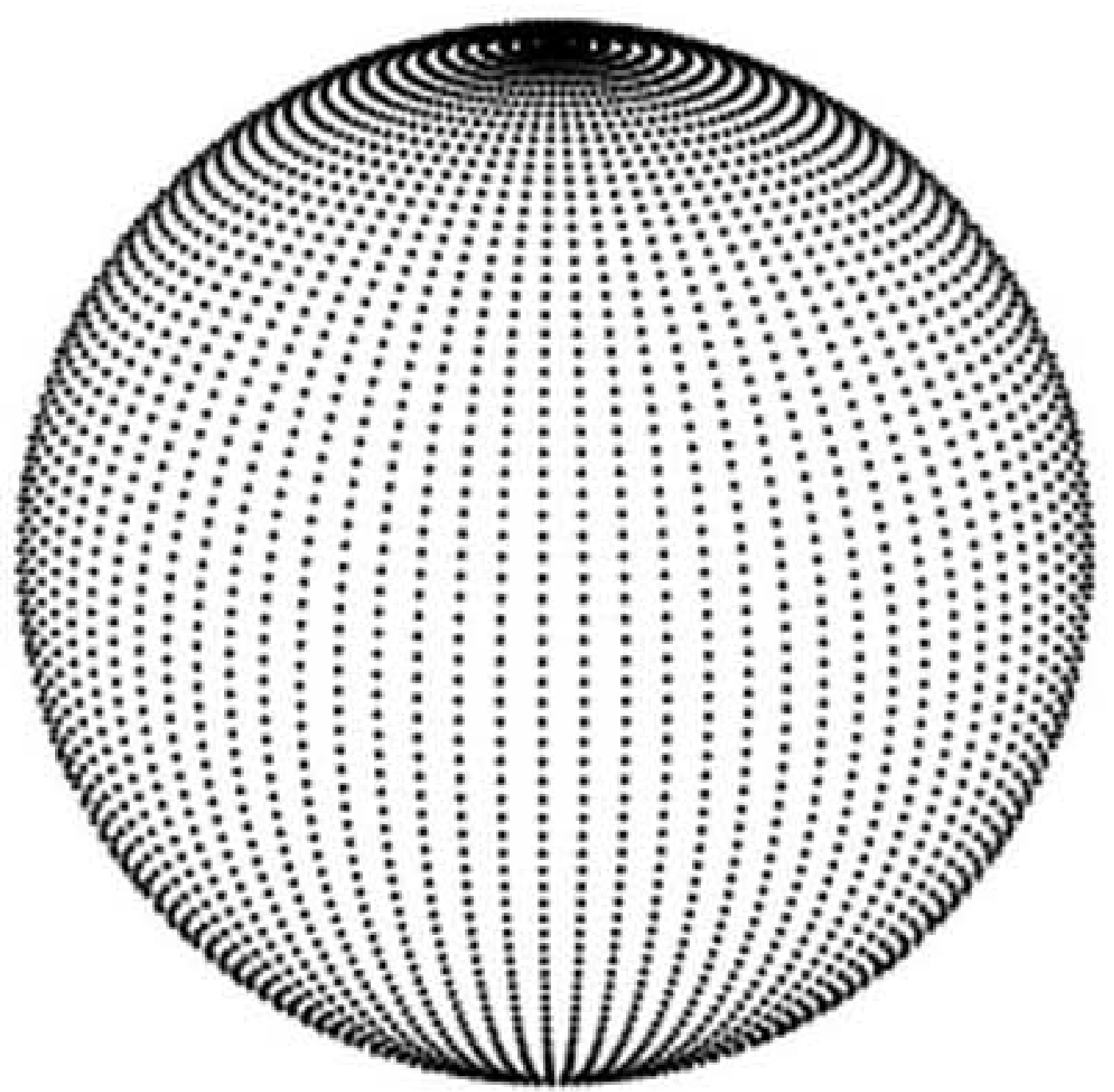}&\hspace{0.5cm}\includegraphics[width=3.7cm]{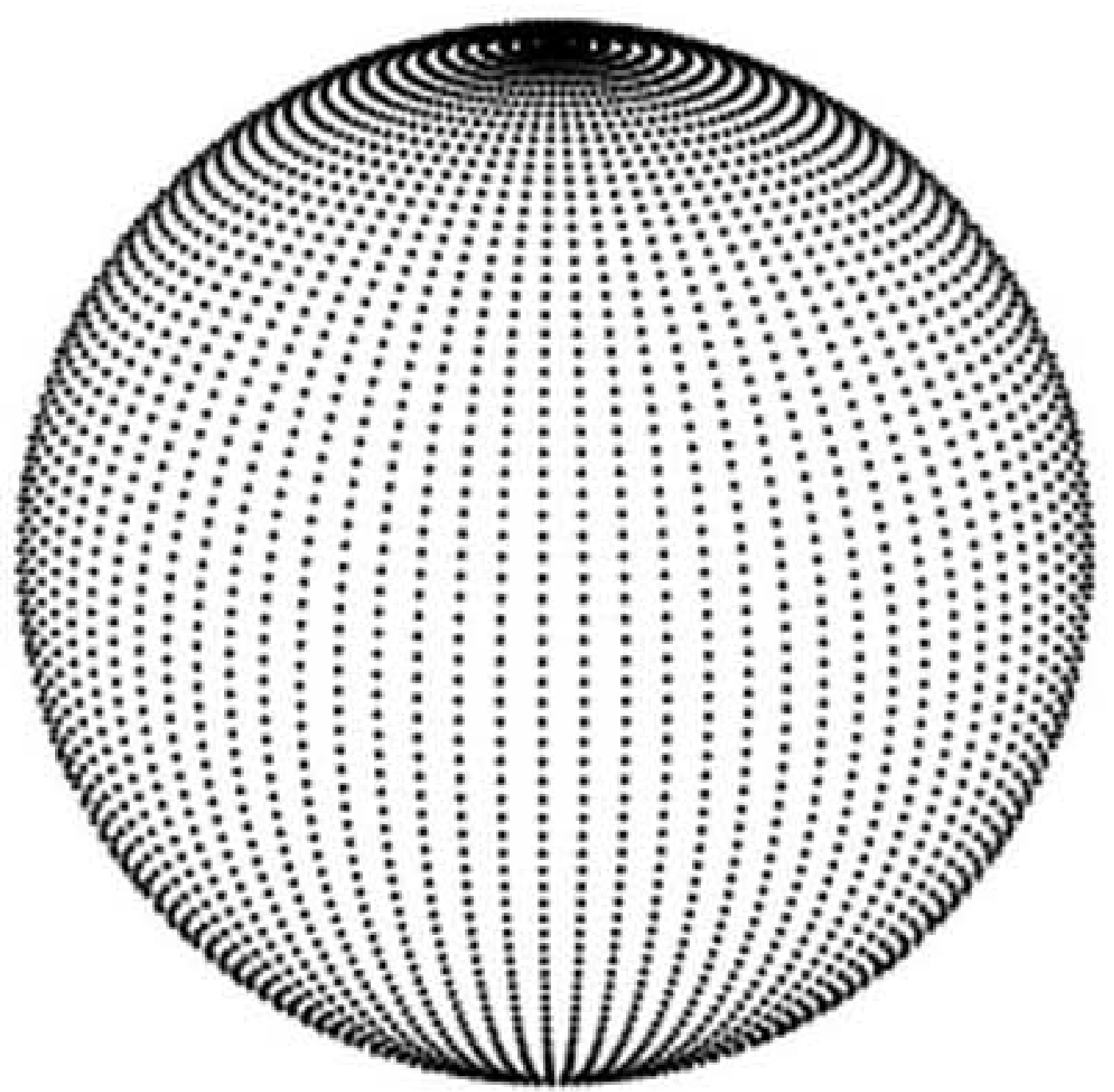}\\
\end{tabular}
\caption{Upper panel: Spot migration diagram for the first spot of the secondary of KIC~8553788 during Q14. Lower panels: The location of the spot (crosses) on the surface of the secondary component during the first day (left) and the last day (right), when the system is at orbital phase 0.24.}
\label{fig:spotmigr2a}

\begin{tabular}{cc}
\multicolumn{2}{c}{\includegraphics[width=7.5cm]{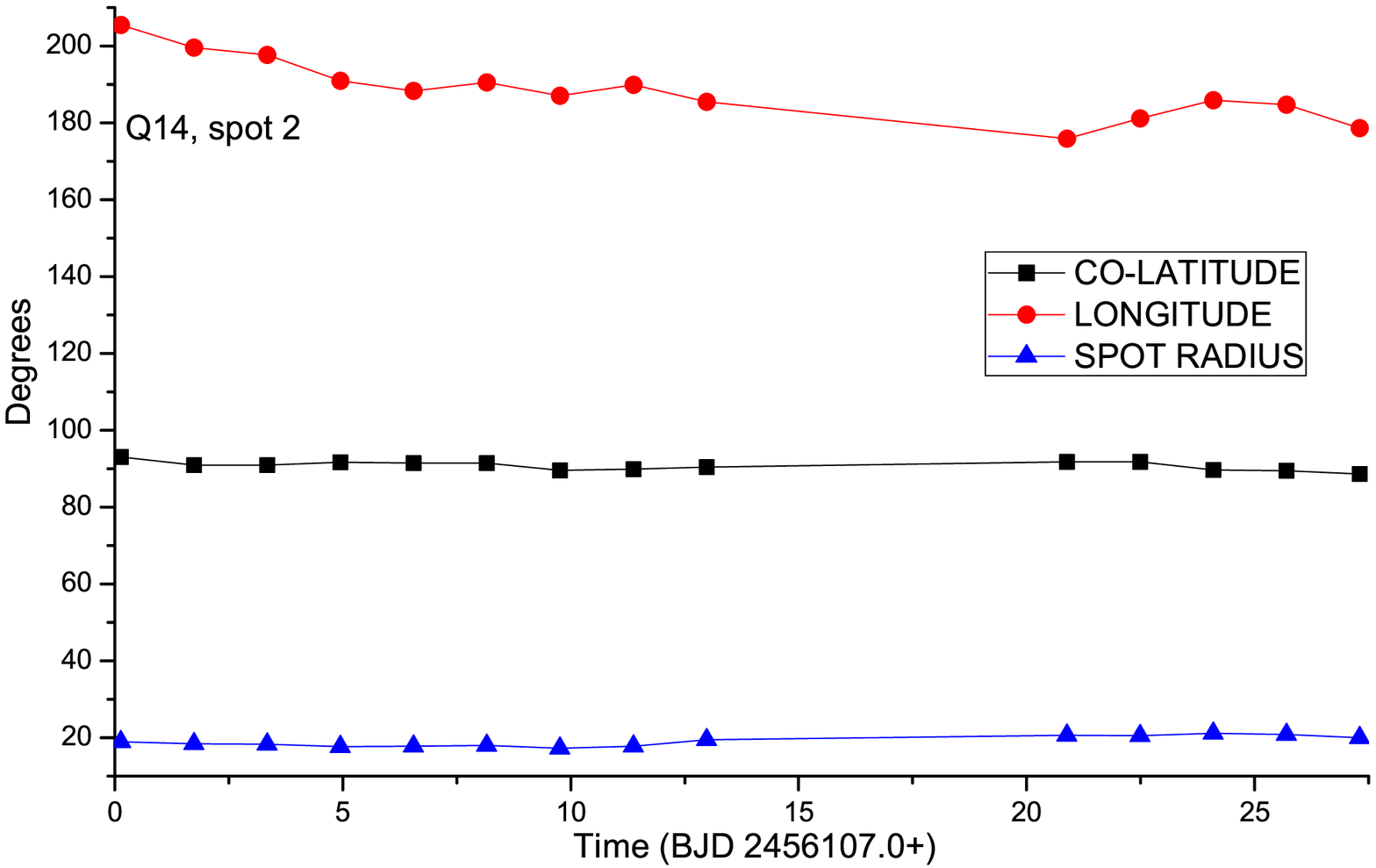}}\\
\includegraphics[width=3.7cm]{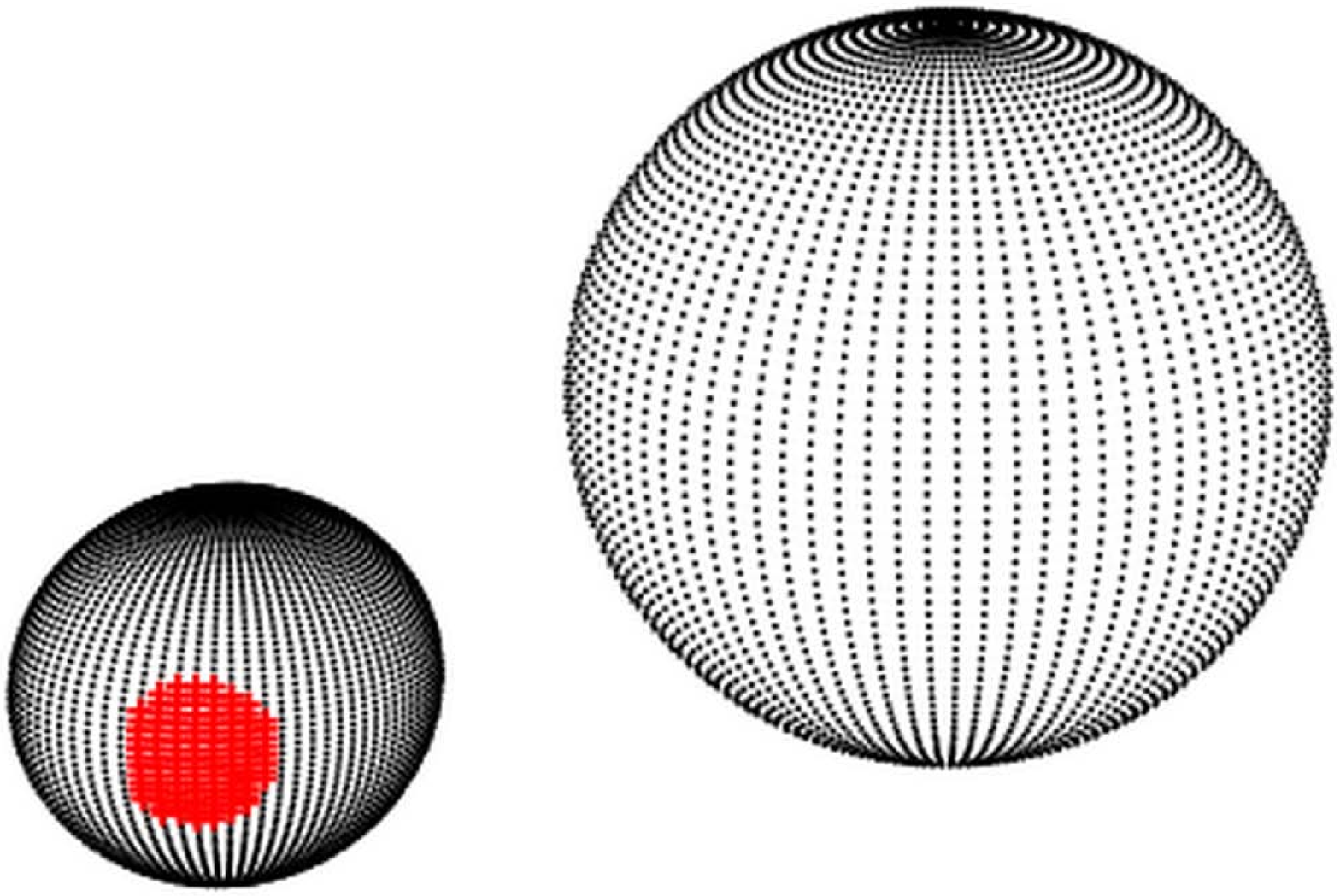}&\hspace{0.5cm}\includegraphics[width=3.7cm]{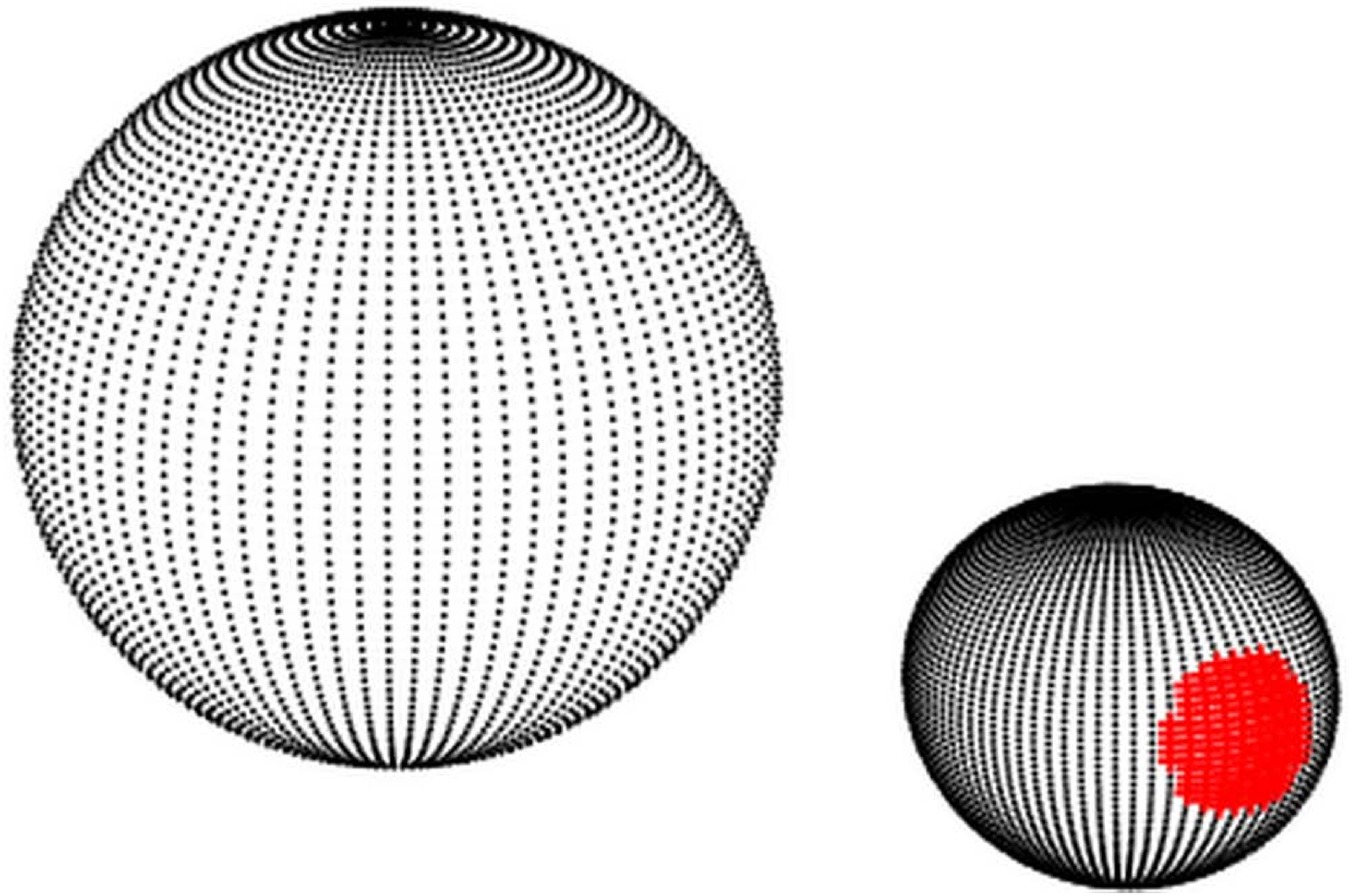}\\
\end{tabular}
\caption{The same as Fig.~\ref{fig:spotmigr2a}, but for the second spot of KIC~8553788, when the system is at orbital phase 0.91 (lower left) and at 0.09 (lower right).}
\label{fig:spotmigr2b}
\end{figure}

\end{appendix}

\end{document}